\documentclass[12pt]{article}
\usepackage{latexsym,amsmath,amssymb,graphicx}
\renewcommand{\thefootnote}{\fnsymbol{footnote}}

\usepackage{amsmath}
\usepackage{amsfonts}

\numberwithin{equation}{section}

\def\doubleset#1#2{\bgroup%
\def\doit#1#2{%
\setbox\dblsetbox=\hbox{$\cstyle #1$}%
\raise#2\ht\dblsetbox\copy\dblsetbox%
\hskip-\wd\dblsetbox%
\raise-#2\ht\dblsetbox\box\dblsetbox}%
\mathchoice%
{\def\cstyle{\displaystyle}\doit#1#2}%
{\def\cstyle{\textstyle}\doit#1#2}%
{\def\cstyle{\scriptstyle}\doit#1#2}%
{\def\cstyle{\scriptscriptstyle}\doit#1#2}\egroup}
\def\underarrow#1{\vbox{\ialign{##\crcr$\hfil\displaystyle
 {#1}\hfil$\crcr\noalign{\kern1pt\nointerlineskip}$\longrightarrow$\crcr}}}

\newbox\dblsetbox

\newlength{\extraspace}
\newlength{\extraspaces}
\newcommand{\be}{\begin{equation}
\addtolength{\abovedisplayskip}{\extraspaces}
\addtolength{\belowdisplayskip}{\extraspaces}
\addtolength{\abovedisplayshortskip}{\extraspace}
\addtolength{\belowdisplayshortskip}{\extraspace}}
\newcommand{\ee}{\end{equation}}

\newcommand{\ba}{\begin{eqnarray}
\addtolength{\abovedisplayskip}{\extraspaces}
\addtolength{\belowdisplayskip}{\extraspaces}
\addtolength{\abovedisplayshortskip}{\extraspace}
\addtolength{\belowdisplayshortskip}{\extraspace}}
\newcommand{\ea}{\end{eqnarray}}

\newcommand{\bd}{\begin{displaymath}
\addtolength{\abovedisplayskip}{\extraspaces}
\addtolength{\belowdisplayskip}{\extraspaces}
\addtolength{\abovedisplayshortskip}{\extraspace}
\addtolength{\belowdisplayshortskip}{\extraspace}}
\newcommand{\ed}{\end{displaymath}}

\newcounter{saveeqn}

\newcommand{\newsection}[1]{
\vspace{12mm} \pagebreak[3] \addtocounter{section}{1}
\setcounter{equation}{0} \setcounter{subsection}{0}
\noindent{\bf \thesection. #1} \nopagebreak
\medskip
\nopagebreak}

\newcommand{\newsubsection}[1]{
\vspace{0.8cm} \pagebreak[3] \addtocounter{subsection}{1}
\noindent{\it \thesubsection. #1} \nopagebreak \vspace{2mm}
\nopagebreak}

\begin{document}
\addtolength{\baselineskip}{1.5mm}

\thispagestyle{empty}

\vbox{} \vspace{1.5cm}

\begin{center}
\centerline{\LARGE{Five-Branes In M-Theory}}
\bigskip
\centerline{\LARGE{And A Two-Dimensional  Geometric Langlands Duality}}
\bigskip
\centerline{\LARGE{}}

\vspace{1.5cm}

{Meng-Chwan~Tan \footnote{On leave of absence from the National
University of Singapore.}}
\\[2mm]
{\it School of Natural Sciences, Institute for Advanced Study\\
Princeton, New Jersey 08540, USA} \\[1mm]
e-mail: tan@ias.edu\\
\end{center}

\vspace{1.5 cm}

\centerline{\bf Abstract}\smallskip \noindent
 A recent attempt to extend the geometric Langlands duality to affine Kac-Moody groups, has led Braverman and Finkelberg~\cite{BF} to conjecture a mathematical relation between the intersection cohomology of the moduli space of $G$-bundles on certain singular complex $\it{surfaces}$, and the integrable representations of the $\it{Langlands}$ $\it{dual}$ of an associated affine $G$-algebra, where $G$ is any simply-connected semisimple group. For the $A_{N-1}$ groups, where the conjecture has been mathematically verified to a large extent, we show that the relation has a natural physical interpretation in terms of six-dimensional compactifications of M-theory with coincident five-branes wrapping certain hyperk\"ahler four-manifolds; in particular, it can be understood as an expected invariance in the resulting spacetime BPS spectrum under string dualities. By replacing the singular complex surface with a $\it{smooth}$ multi-Taub-NUT manifold, we find agreement with a closely related result demonstrated earlier via purely field-theoretic considerations by Witten~\cite{Lectures by Witten}. By adding OM five-planes to the original analysis, we show that the relation ought to hold for the $\textrm{\it{non-simply-connected}}$ $D_N$ groups as well. This is the first example of a string-theoretic interpretation of such a two-dimensional extension to complex surfaces of the geometric Langlands duality for the $A-D$ groups.

 \newpage

\renewcommand{\thefootnote}{\arabic{footnote}}
\setcounter{footnote}{0}

\newsection{Introduction}

  A recent attempt to extend the geometric Langlands duality to affine Kac-Moody groups was undertaken by Braverman and Finkelberg in \cite{BF}. Their approach towards such an extension was to find an analog of the geometric Satake isomorphism - which underlies the original geometric Langlands duality - for affine Kac-Moody groups. In doing so, they arrived at a conjectured relation between the intersection cohomology of the moduli space of $G$-bundles on the $\it{singular}$ ALE space ${\mathbb R}^4/{\mathbb Z}_{k}$, and the integrable representations of the $\it{Langlands}$ $\it{dual}$ of an associated affine $G$-algebra at level $k$, where $G$ is any semisimple group that is simply-connected. In contrast to the original geometric Langlands conjecture, this relation involves the moduli space of $G$-bundles over (singular) complex surfaces instead of curves. Hence, it can, in this sense, be regarded as a definition for a $\it{two}$-$\it{dimensional}$ extension to complex surfaces of the geometric Langlands duality. The case where $G$ is of $A_{N-1}$ type was proved to a large extent in \cite{BF}. However, it remains an outstanding task to prove the conjecture for all other semisimple groups that are simply-connected.

In this paper, we furnish a purely physical interpretation of the above-mentioned Braverman-Finkelberg relation for $A_{N-1}$ groups; in particular, we will show that the relation has a natural physical interpretation in terms of six-dimensional compactifications of M-theory with coincident five-branes wrapping certain hyperk\"ahler four-manifolds. Specifically, it can be understood as an expected invariance in the resulting spacetime BPS spectrum under string dualities. Our findings thus serve to provide a physical corroboration of the proof of the relation for $A_{N-1}$ groups. On the other hand, the consistency of our physical arguments with this $\it{a~priori}$ abstract relation, yet again lends support to the mathematical validity of dualities in string theory. By replacing the singular complex surface $\mathbb R^4/\mathbb Z_k$ with the $\it{smooth}$ multi-Taub-NUT manifold with $k$ centres, we make contact and find agreement with a closely related result - involving a relation between the ${\bf L}^2$-cohomology of the moduli space of $G$-instantons on the multi-Taub-NUT manifold, and the tensor product of $k$ representations of the corresponding loop group $\mathcal{L} G$ at level 1 - demonstrated earlier via purely field-theoretic considerations by Witten~\cite{Lectures by Witten}. In addition, we also discuss how this result can possibly be interpreted as a generalisation of Nakajima's celebrated result in \cite{naka}. Last but not least, by adding OM five-planes to the original analysis on $\mathbb R^4/\mathbb Z_k$, we argue that the Braverman-Finkelberg relation ought to hold for the $\textrm{\it{non-simply-connected}}$ $D_N$ groups as well. This is the first example of a string-theoretic interpretation of such an extension of the well-known geometric Langlands duality for the $A-D$ groups.

We shall now give a brief summary and plan of the paper. In $\S$2, we will review, in preparation for the discussion in $\S$3, the hyperk\"ahler multi-Taub-NUT space and a certain hyperk\"ahler four-manifold discussed by Sen in \cite{sen}. We will also discuss their significance in type IIA compactifications of M-theory and T-dualities in string theory. In $\S$3, we will introduce a chain of dualities which will allow us to relate two distinct but $\it{dual}$ six-dimensional compactifications of M-theory with coincident five-branes (and OM five-planes) wrapping the complex $\mathbb R^4/\mathbb Z_k$ surface and the above-mentioned hyperk\"ahler four-manifold discussed by Sen. In $\S$4, we will first review some essential facts about the Braverman-Finkelberg conjecture and the relation it implies. Then, we will proceed to explain the physical interpretation of the relation for $A_{N-1}$ groups. Next, we will replace the complex $\mathbb R^4/\mathbb Z_k$ surface with the smooth multi-Taub-NUT manifold with $k$ centres, and derive Witten's earlier result from a purely string-theoretic perspective. Finally in $\S$5, we will reconsider our original analysis on $\mathbb R^4/\mathbb Z_k$ in the presence of OM five-planes, and argue in favour of a Braverman-Finkelberg type relation for the non-simply-connected $D_N$ groups.

This paper grew out of an attempt to better understand a series of lectures delivered by E.~Witten at the IAS entitled ``Duality From Six-Dimensions'' in Feb 08, from which the initial insights for the present work was also gained. I would first and foremost like to thank him for providing all the answers to my questions in regard to his lectures and more. I have also benefitted greatly from many a discussion with A.~Braverman and H.~Nakajima. My deepest gratitude goes out to them for their patience and time in educating me on their work and related matters.  Last but not least, I would like to thank A.~Basu, O.~Bergman, S.A.~Cherkis, J.~Fuchs, V.~Pestun, A.~Sen, Y.~Tachikawa and D.~Tong, for highly illuminating exchanges.

This work is supported by the Institute for Advanced Study and the NUS-Overseas Postdoctoral Fellowship.

\newsection{The Multi-Taub-NUT Space, Sen's Four-Manifold and String/M-Theory}

In this section, we will review the multi-Taub-NUT space and Sen's four-manifold in the context of type IIA compactifications of M-theory and T-dualities in string theory. We shall begin by discussing aspects of their geometries which will be relevant to our later discussions, and then explain how these four-manifolds can be associated with D6-branes, NS5-branes, and ON$5^-$-planes in type IIA and IIB string theories. Careful attention will be paid to various subtleties that are rarely emphasised in the physics literature but are nonetheless central to our arguments in the next section. The reader who is absolutely familiar with such matters can skip this section if desired.

\newsubsection{The Geometry of Multi-Taub-NUT Space}

The multi-Taub-NUT space, which we will henceforth denote broadly as $TN_k$, is a hyperk\"ahler four-manifold that can be regarded as a non-trivial, singular ${\bf S}^1$ fibration over ${\mathbb R}^3$. It has the metric~\cite{BBS}
\be
ds^2_{TN_k} = {1\over U(\vec{r})}(d\alpha + \chi)^2 + U(\vec{r})d\vec{r}^2,
\label{TN}
\ee
where $\alpha$ is a compact periodic coordinate, and ${\vec r} = (r^1, r^2, r^3)$ is a three-vector in $\mathbb R^3$. The function $U(\vec{r})$ and the 1-form $\chi$ are defined by
\be
U(\vec{r}) = 1 + {R\over 2}\sum_{a=1}^k {1 \over{|\vec{r} - \vec {r}_a}|}, \qquad\qquad  d\chi = *_3~dU,
\label{TN function}
\ee
where $*_3$ is Poincar\'e duality in three-dimensions. Smoothness requirements of the metric (\ref{TN}) dictate that $\alpha$ must have period $2 \pi R$. Hence, the actual radius of the circle fibre is given by~\cite{BBS}
\be
{\widetilde R}(\vec{r}) = U(\vec{r}) ^{-1/2} R.
\label{radius}
\ee

Now, notice from (\ref{TN function}) and (\ref{radius}) that the circle fibre shrinks to zero size at the $k$ points $\vec{r}_1, \vec{r}_2, \dots, \vec{r}_k$ in $\mathbb R^3$. That is, there are $k-1$ line segments that connect each pair of neighboring points, and over each of these $k-1$ line segments, there is a circle fibration which degenerates at the end points. In other words, $TN_k$ is generically a perfectly smooth four-manifold with $k-1$ homologically independent two-spheres given by the circle fibrations over the line segments.

In addition, also notice from (\ref{TN function}) and (\ref{radius}) that at infinity, i.e. $\vec{r} \to \infty$, we have ${\widetilde R}(\infty) = R$. Consequently, one can see from (\ref{TN}) that the geometry of $TN_k$ at infinity approximates $\mathbb R^3 \times {{\bf S}^1}$, where ${\bf S}^1$ has a fixed radius of $R$. However, the ${\bf S}^1$ factor is actually non-trivially fibred over the ${\bf S}^2 $ submanifold of $\mathbb R^3 \cong {\bf S}^2 \times \mathbb R$ at infinity, where the fibration can be viewed as a monopole bundle of charge (or first Chern-class) $k$, i.e.,
\be
\int_{{\bf S}^2} d\chi = 2\pi k.
\label{monopole charge}
\ee
This point will be important when we discuss $TN_k$ as an M-theory background and its interpretation as D6-branes in the corresponding type IIA theory.

Next, note that as we `decompactify' the asymptotic radius of the circle by letting $R \to \infty$, the geometry of $TN_k$ will be that of a $\it{resolved}$ ALE space of type $A_{k-1}$; the intersection matrix of the two-spheres just gives the Cartan matrix of the $A_{k-1}$ Lie algebra. In order to obtain a $\it{singular}$ ALE space of type $A_{k-1}$ such as $\mathbb R^4 / \mathbb Z_k$, one just needs to bring together all the $k$ points $\vec{r}_1, \vec{r}_2, \dots, \vec{r}_k$ to the origin in $\mathbb R^3$, such that the $k-1$ homologically independent two-spheres all collapse to result in an $A_{k-1}$ singularity at $0$. This has an interpretation in terms of enhanced gauge symmetries in the context of string/M-theory as we will soon explain.


\newsubsection{The Multi-Taub-NUT Space in a IIA/M-Theory Correspondence}

The $k$ Kaluza-Klein monopoles solution in M-theory can be described by the metric
\be
ds^2 = -dt^2 + \sum_{m=1}^{6} dy^m dy^m + ds^2_{TN_k},
\label{monopole solution}
\ee
where the $y^m$'s denote the space-like worldvolume coordinates on the six-dimensional solitons in type IIA that are represented by the above solution in M-theory. In order to ascertain what these solitons are, let us take the ``eleventh circle'' to be the circle fibre of $TN_k$. Then, a D0-brane in type IIA can be interpreted as a Kaluza-Klein excitation along the ``eleventh circle''. The D0-brane is electrically charged under the gauge field $C_\mu = g_{\mu 11}$ after a Kaluza-Klein reduction. Therefore, its magnetic dual, the D6-brane, must be magnetically charged under the same gauge field. Since a Kaluza-Klein monopole must correspond to a magnetically charged soliton, we find that the six-dimensional space with coordinates $y^m$ ought to be filled by D6-branes after a type IIA compactification of M-theory along the circle fibre of $TN_k$.

That one has $k$ D6-branes is consistent with  the fact that the circle fibration of $TN_k$ is actually a monopole bundle of charge $k$ at infinity via (\ref{monopole charge}). Note also that the $\vec{r}_a$'s can be interpreted as the location of the Kaluza-Klein monopoles in $\mathbb R^3 \in TN_k$. This means that the $k$ D6-branes will be localised at the $k$ points $\vec{r}_1, \vec{r}_2, \dots, \vec{r}_k$ in $\mathbb R^3 \in TN_k$. Therefore, as one brings the $k$ points together towards $0$, all $k$ D6-branes will coincide and the worldvolume theory will possess an enhanced non-abelian $U(k)$ gauge symmetry. Hence, upon a compactification along the circle fibre of M-theory on a $TN_k$ that has an $A_{k-1}$ singularity at its origin, one will obtain an equivalent description in terms of a stack of $k$ $\it{coincident}$ D6-branes that span the directions transverse to $TN_k$ in type IIA string theory. One can also understand this enhancement of gauge symmetries as follows~\cite{sen}. Starting with a non-singular $TN_k$ manifold, there are M2-branes which wrap the $k-1$ two spheres in $TN_k$. Upon compactification along the circle fibre, these M2-branes become open strings in type IIA which connect between neighboring D6-branes which are non-coincident. As we bring all the $\vec{r}_a$'s together, the $k-1$ two-spheres in $TN_k$ collapse, and we have an enhanced gauge symmetry in M-theory due to extra massless gauge fields that originate from the M2-branes which now have zero-volume, in the transverse spacetime directions. In the equivalent IIA picture, this corresponds to the open strings becoming massless as the $k$ D6-branes become coincident, which consequently results in an enhanced non-abelian gauge symmetry in the transverse spacetime directions along the worldvolume of the D6-branes.

Another relevant point would be the following. In order for the tension of a soliton described by the monopole solution (\ref{monopole solution}) to agree with the tension of a D6-brane in type IIA string theory, one has to set $R = g^A_s l_s$, where $g^A_s$ is the IIA string coupling and $l_s$ is the string length scale~\cite{BBS}. In particular, a compactification of M-theory along the circle fibre of $TN_k$ where the asymptotic radius $R$ is either large or small, will result in an equivalent IIA theory that is either strongly or weakly coupled, respectively.

\newsubsection{The Multi-Taub-NUT Space, NS5-Branes and T-Duality}

Let us now consider the following ten-dimensional background in type IIA or IIB string theory:
\be
ds^2 = -dt^2 + \sum_{l=1}^{5} dy^l dy^l + ds^2_{TN_k}.
\label{type II background}
\ee
Notice that the metric (\ref{TN}) enjoys a $U(1)$ isometry which acts to shift the value of $\alpha$. Consequently, this allows for the application of T-duality transformations to the above background solution. In doing so, one will obtain the following T-dual solution~\cite{Tong, CJ}:
\be
ds^2 = -dt^2 + \sum_{l=1}^{5} dy^l dy^l + V(\vec{x})(d\theta^2 + d\vec{r}^2),
\label{type II background-T-dual}
\ee
where $\theta$ is a compact coordinate of period $2\pi$ which parameterises the dual ${\bf S}^1$, and
\be
V(\vec{x}) = {1\over R^2} + {1\over 2} \sum_{a=1}^{k} {1\over |\vec{x} - \vec{x}_a|},
\label{V}
\ee
where $\vec{x} = (\theta, \vec{r})$ is taken to mean a position in a full $\mathbb R^4$. From (\ref{V}) and (\ref{type II background-T-dual}), we see that the asymptotic radius of the dual circle is indeed given by $1/R$ as expected under T-duality.

Note that the solution given by (\ref{type II background-T-dual}) consists of $k$ objects which are pointlike in the $\mathbb R^4$, and which are also magnetic sources of the NS-NS potential $B_{\mu \nu}$~\cite{CJ}. In fact, they just correspond to $k$ NS5-branes which span the space with coordinates $y^l$, that are also arranged in a circle on $\theta$ and localised on the rest of $\mathbb R^4$ according to the centres $\vec{x}_a, a = 1, 2, \dots, k$. Reversing the above arguments, we conclude that one can do a T-duality along any circle that is $\it{transverse}$ to a stack of $k$ $\it{coincident}$ NS5-branes in type IIA(IIB) string theory, and obtain a dual background with no NS5-branes but with a $TN_k$ manifold that has an $A_{k-1}$ $\it{singularity}$ at the origin in type IIB(IIA) string theory. In addition, notice that the asymptotic radius $R$ of the dual, singular $TN_k$ background must tend to zero if the radius ${V(\vec{x})}^{1/2}$ of the circle transverse to the NS5-branes is to be infinitely large at any point $\vec{r} \in \mathbb R^3$.

Last but not least, note that in going from (\ref{type II background}) to (\ref{type II background-T-dual}) under  T-duality transformations, only components of the solution transverse to the NS5-brane worldvolume get modified. In other words, the components of the solution along the worldvolume directions have no structure and are therefore trivial. Consequently, an application of T-duality along any worldvolume direction will map us back to the same NS5-brane solution given by (\ref{type II background-T-dual}).\footnote{This is to be contrasted with a $D_p$-brane, where T-duality along a direction parallel or transverse to its worldvolume will result in a $D_{p-1}$ or $D_{p+1}$-brane, respectively.}

\newsubsection{The Geometry of Sen's Four-Manifold}

Let us consider the following four-manifold characterised by a non-trivial ${\bf S}^1$ fibration over $\mathbb R^3$ with metric~\cite{sen}
\be
ds^2 = {1 \over W(\vec{r})} (d\alpha + \chi)^2 + W(\vec{r}) d\vec{r}^2,
\label{sen's manifold}
\ee
modded out by the transformation
\be
(\vec{r} \rightarrow -\vec{r}, \qquad \alpha \rightarrow -\alpha),
\label{reflection}
\ee
where $\alpha$ is a compact periodic coordinate of the ${\bf S}^1$ fibre, and $\vec{r} = (r^1, r^2, r^3)$ is a three-vector in $\mathbb R^3$. The function $W(\vec{r})$ and the 1-form $\chi$ are defined by
\be
W(\vec{r}) = 1 - {2R \over {|\vec{r}|} }  + {R\over 2}\sum_{a=1}^k \left( {1\over |\vec{r} - \vec{r}_a|} + {1\over |\vec{r} + \vec{r}_a|} \right), \qquad d\chi = *_3~dW,
\label{SN function}
\ee
where $*_3$ is Poincar\'e duality in three-dimensions, and where the asymptotic radius of the circle fibre is $R$ (before the identification in (\ref{reflection})).

Note that the metric is invariant under the reflection (\ref{reflection}); $W(\vec{r})$ is invariant under $(\vec{r} \rightarrow -\vec{r})$ and $\chi$ changes sign under the reflection. However, the metric is singular at $\vec{r} =0$. This singularity can be removed by replacing the metric near $\vec{r} =0$ by the Atiyah-Hitchin metric~\cite{sen18}, which is completely non-singular after we perform the reflection (\ref{reflection}). We shall henceforth denote this effectively smooth, hyperk\"ahler four-manifold broadly as Sen's four-manifold or $SN_k$.

In the region where $\vec{r} \to \infty$, we see from (\ref{SN function}) that $W(\vec{r}) \to 1$. Hence, from (\ref{sen's manifold}) and (\ref{reflection}), we find that $SN_k$ approximates $(\mathbb R^3 \times {{\bf S}^1})/{\cal I}_4$ far away from the origin at infinity, where ${\cal I}_4$ denotes an independent action on the two factors $\mathbb R^3$ and ${\bf S}^1$ that is defined in (\ref{reflection}).  As mentioned earlier, the ${\bf S}^1$ factor has a fixed radius of $R$.

At the $k$ points $\vec{r}_1, \dots, \vec{r}_k$ in $SN_k$, the circle fibre shrinks to zero size, as one can see from (\ref{SN function}) and (\ref{sen's manifold}). Consequently, the circle fibrations over the line segments connecting each of these neighboring points will result in a set of $k-1$ two-spheres. Because the reflection (\ref{reflection}) is a symmetry of the space, there is an identification $\vec{r}_a \sim - \vec{r}_a$. As such, there will be additional two spheres coming from the extra circle fibrations over the line segments that connect the points $\vec{r}_i$ and $-\vec{r}_{i+1}$. In short, the homologically independent two-spheres will define an intersection matrix that is the Cartan matrix of a $D_k$ Lie algebra~\cite{sen}. If we let all the $\vec{r}_a$'s approach the origin, the areas of all the two-spheres vanish, and we obtain a $D_k$ singularity. As we shall explain below, this observation is consistent with the fact that such an $SN_k$ background in string/M-theory would lead to an enhanced $SO(2k)$ gauge symmetry.


\newsubsection{Sen's Four-Manifold in a IIA/M-Theory Correspondence}

Consider the following eleven-dimensional background in M-theory:
\be
ds^2 = -dt^2 + \sum_{m=1}^{6} dy^m dy^m + ds^2_{SN_k},
\label{orientifold solution}
\ee
where the $y^m$'s denote the space-like worldvolume coordinates on the six-dimensional solitons in type IIA that are represented by the above solution in M-theory. In order to ascertain what these solitons are, first note that near $\vec{r}=0$, the metric of $SN_k$ agrees with the Atiyah-Hitchin or AH space. It is known that upon a type IIA compactifcation of M-theory along the circle fibre of such an AH space, one would get an orientifold six-plane~\cite{sen17}. Second, note that near the point $\vec{r} = \vec{r}_a$ or its image $-\vec{r}_a$ (under ${\cal I}_4$) for $1\leq a \leq k$, the metric agrees with the one near a Kaluza-Klein monopole. Moreover, far away from the origin at infinity, the metric looks like the multi-Taub-NUT space at infinity albeit identified under the action of ${\cal I}_4$. In all, this means that (\ref{orientifold solution}) represents an M-theory background which upon compactification along the circle fibre, gives us $k$ D6-branes and an O6-plane in type IIA string theory which span the directions transverse to $SN_k$ given by the coordinates $y^m$.\footnote{As emphasised in \cite{sen} itself, the M-theory background given by (\ref{orientifold solution}) is only an approximate solution to the exact one describing the D6-branes and O6-plane in type IIA string theory. However, it differs from the exact solution by terms that vanish exponentially as we move away from the origin. Since our penultimate discussion in $\S$5 will only involve an analysis of $SN_k$ near the boundary at infinity, this deviation from the exact solution will not affect us.}

Note also that the $\vec{r}_a$'s can be interpreted as the location of the Kaluza-Klein monopoles in $SN_k$. This means that the $k$ D6-branes will be localised at the $k$ points $\vec{r}_1, \vec{r}_2, \dots, \vec{r}_k$ in $SN_k$. Therefore, as one brings the $k$ points together towards $0$, all $k$ D6-branes will coincide on top of the O6-plane and the worldvolume theory will possess an enhanced non-abelian $SO(2k)$ gauge symmetry.\footnote{One has an $SO(2k)$ gauge symmetry because of the presence of an O6-plane. Here and henceforth, an O6-plane will mean an $\textrm{O}6^{-}$ plane, i.e., the orientifold six-plane that is associated with a worldsheet parity operator whose eigenvalue is $-1$.} Hence, upon a compactification along the circle fibre of M-theory on an $SN_k$ that has a $D_k$ singularity at its origin, one will obtain an equivalent description in terms of a stack of $k$ $\it{coincident}$ D6-branes on top of an O6-plane that span the directions transverse to $SN_k$ in type IIA string theory. One can also understand this enhancement of gauge symmetries from the perspective of M2-branes wrapping the two-spheres in $SN_k$ and open strings in type IIA connecting between the D6-branes~\cite{sen}. Since the discussion is analogous to the one before on $TN_k$, we shall skip it for brevity.

Once again, in order for the tension of a soliton described by the monopole solution (\ref{orientifold solution}) to agree with the tension of a D6-brane in type IIA string theory, one must have $R \sim g^A_s l_s$. Therefore, a compactification of M-theory along the circle fibre of $SN_k$ where the asymptotic radius $R$ is either large or small, will result in an equivalent IIA theory that is either strongly or weakly coupled, respectively.

\newsubsection{Sen's Four-Manifold, NS5-branes/ON5-planes and T-Duality}

Consider the following ten-dimensional background in either type IIA or IIB string theory:
\be
ds^2 = -dt^2 + \sum_{l=1}^{5} dy^l dy^l + ds^2_{SN_k}.
\label{type II background Sen}
\ee
Notice that the metric (\ref{sen's manifold}), just like the metric (\ref{TN}), enjoys a $U(1)$ isometry which acts to shift the value of $\alpha$. Consequently, this allows for the application of T-duality transformations to the above background solution, just like in the multi-Taub-NUT example.  Far away from the origin,\footnote{As mentioned earlier, our main analysis in $\S$5 will only involve the physics of the background near infinity. As such, it suffices to discuss what happens away from the origin only.} the T-dual background will therefore look like
\be
ds^2 = -dt^2 + \sum_{l=1}^{5} dy^l dy^l + Y(\vec{r})(d\theta^2 + d\vec{r}^2),
\label{type II background-T-dual Sen}
\ee
where $\theta$ is a compact coordinate of period $2\pi$ which parameterises the dual ${\bf S}^1$, and
\be
Y(\vec{x}) = {1\over R^2} - {2 \over {|\vec{x}|}}  + {1\over 2}\sum_{a=1}^k \left( {1\over |\vec{x} - \vec{x}_a|} + {1\over |\vec{x} + \vec{x}_a|} \right),
\label{Y}
\ee
where $\vec{x} = (\theta, \vec{r})$ is taken to mean a position in a full $\mathbb R^4$. From (\ref{Y}) and (\ref{type II background-T-dual Sen}), we see that the asymptotic radius of the dual circle is indeed given by $1/R$ as expected under T-duality.

Note that the solution given by (\ref{type II background-T-dual Sen}) consists of $2k$ objects which are pointlike in the $\mathbb R^4$, and which are also magnetic sources of the NS-NS potential $B_{\mu \nu}$~\cite{CJ}. In fact, they just correspond to $2k$ NS5-branes which span the space with coordinates $y^l$, that are localised on the $\mathbb R^4$ according to the centres $\pm\vec{x}_a, a = 1, 2, \dots, k$. The reason why we ended up with a dual background that appears to have $2k$ instead of $k$ NS5-branes is because the background represented by (\ref{sen's manifold})-(\ref{SN function}), and therefore the type II background (\ref{type II background Sen}), incorporates a reflection in the spatial directions transverse to the NS5-branes, which, effectively doubles the number of NS5-branes present. This means that the T-dual solution (\ref{type II background-T-dual Sen}) really corresponds to a background which only has $k$ dynamical NS5-branes and an ON$5^-$-plane, whereby the `-' superscript just indicates that its presence will result in an orthogonal gauge symmetry in the worldvolume theory as required, while the `N' just denotes that it can only be associated with NS5-branes~\cite{hanany}. Reversing the above arguments, we conclude that one can do a T-duality along any circle that is $\it{transverse}$ to a stack of $k$ $\it{coincident}$ NS5-branes on top of an ON$5^-$-plane in type IIA(IIB) string theory, and obtain a dual background with no NS5-branes and no ON$5^-$-plane but with an $SN_k$ manifold that has a $D_k$ $\it{singularity}$ at the origin in type (IIB)(IIA) string theory. In addition, notice that the asymptotic radius $R$ of the dual, singular $SN_k$ background must tend to zero if the radius ${Y(\vec{x})}^{1/2}$ of the circle transverse to the NS5-branes is to be infinitely large over any point $\vec{r} \in \mathbb R^3$.

Last but not least, note that in going from (\ref{type II background Sen}) to (\ref{type II background-T-dual Sen}) under  T-duality transformations, only components of the solution transverse to the NS5-brane/ON$5^-$-plane worldvolume get modified. In other words, the components of the solution along the worldvolume directions have no structure and are therefore trivial. Consequently, an application of T-duality along any worldvolume direction will map us back to the same NS5-brane/ON$5^-$-plane solution given by (\ref{type II background-T-dual Sen}).

\newsection{A Chain of Dualities}

In this section, we shall introduce a chain of dualities that will allow us to relate two distinct but $\it{dual}$ six-dimensional compactifications of M-theory with stacks of coincident M5-branes wrapping the compactified directions. We will then repeat the arguments with the addition of an OM5-plane to the original stack of coincident M5-branes. In doing so, we will be able to relate a certain six-dimensional M-theoretic compactification with coincident M5-branes and an OM5-plane wrapping the compactified directions, to its dual with $\it{only}$ coincident M5-branes  wrapping the compactified directions. As we shall elucidate in the next two sections, the duality of the first pair of compactifications will provide us with the basis for a physical interpretation of the Braverman-Finkelberg relation for $A_{N-1}$ groups, while the duality of the second pair of compactifications will serve to support our argument for an analogous relation involving the non-simply-connected $D_N$ groups.

\newsubsection{Dual Six-Dimensional Compactifications of M-theory with Five-Branes}

Consider a six-dimensional compactification of M-theory on the five-manifold $\mathbb R^4 / {\mathbb Z_k} \times {{\bf S}^1}$, where $\mathbb R^4/ \mathbb Z_k$ is the singular ALE manifold of type $A_{k-1}$. Wrap on this five-manifold a stack of $N$ coincident M5-branes, such that the worldvolume will be given by $\mathbb R^4 / {\mathbb Z_k} \times {{\bf S}^1} \times \mathbb R_t$, where $\mathbb R_t$ is the time direction. In other words, let us consider the following M-theory configuration (in Euclidean signature):
\be
\textrm{M-theory}: \quad \mathbb R^{5}  \times  \underbrace{\mathbb R^4 / \mathbb Z_k  \times  {{\bf S}^1} \times \mathbb R_t}_{\textrm{$N$ M5-branes}},
\label{M-theory 1}
\ee
Taking the ``eleventh circle'' to be one of the decompactified directions along the $\mathbb R^{5}$ subspace, we see that (\ref{M-theory 1}) actually corresponds to the following ten-dimensional type IIA background with $N$ coincident NS5-branes wrapping $\mathbb R^4 / \mathbb Z_k \times {{\bf S}^1} \times \mathbb R_t$, where $g^A_sl_s \to \infty$:
\be
\textrm{IIA}: \quad \mathbb R^{4}  \times  \underbrace{\mathbb R^4 / \mathbb Z_k  \times  {{\bf S}^1} \times \mathbb R_t}_{\textrm{$N$ NS5-branes}}.
\label{IIA 2}
\ee

 Let us now T-dualise along the $\mathbb R_t$ direction of the worldvolume of the stack of NS5-branes. Recall at this point from our discussion in $\S$2.3, that T-dualising along any one of the worldvolume directions of an NS5-brane (where the solution is trivial), will bring us back to an NS5-brane. Therefore, we will arrive at the following type IIB configuration where $g^B_s \sim 1$ (since $g^B_s = g^A_s l_s /r$, and $r \to \infty$, where $r$ is the radius of ${\mathbb R}_t$):
\be
\textrm{IIB}: \quad \mathbb R^{4}  \times  \underbrace{\mathbb R^4 / \mathbb Z_k  \times  {{\bf S}^1} \times {\mathbb S}^1_{t;  r \to 0}}_{\textrm{$N$ NS5-branes}}.
\label{IIB 3}
\ee

Next, let us T-dualise along a direction that is $\it{transverse}$ to the stack of NS5-branes. As explained in $\S$2.3, one will end up with a $\it{singular}$ $TN_N$ manifold with no NS5-branes. To this end, note that one can view one of the $\mathbb R$'s in $\mathbb R^{4}$ to be a circle of infinite radius. In doing a T-duality along this circle, we arrive at the following type IIA background:
\be
\textrm{IIA}: \quad  TN_N^{R\to 0}  \times  {\mathbb R^4 / \mathbb Z_k}  \times  {{\bf S}^1} \times {\mathbb S}^1_{t; r\to 0},
\label{IIA 4}
\ee
where $TN_N^{R \to 0}$ is a multi-Taub-NUT manifold with an $A_{N-1}$ singularity at the origin and asymptotic radius $R \to 0$. (Recall from $\S$2.3 that $R \to 0$ because we are T-dualising along a trivially fibred circle of infinite radius).  At this stage, one also finds that $g^B_s \to 0$. Consequently, this can be interpreted as the following M-theory background with a very small ``eleventh circle'' ${S}^1_{11}$:
\be
\textrm{M-theory}: \quad {\mathbb S}^1_{t; r \to 0} \times TN_N^{R\to 0} \times {{\bf S}^1} \times S^1_{11; r \to 0} \times {\mathbb R^4 / \mathbb Z_k}.
\label{M-theory 5}
\ee

Recall from $\S$2.1 that the singular ALE space $\mathbb R^4 / \mathbb Z_k$ is simply $TN_k$ with an $A_{k-1}$ singularity at the origin and with asymptotic radius $R \to \infty$. Recall also from $\S$2.2 that M-theory on such a space is equivalent upon compactification along its circle fibre to type IIA string theory with $k$ $\it{coincident}$ D6-branes filling out the directions transverse to the space. In other words, starting from (\ref{M-theory 5}), one can descend back to the following type IIA background:
\be
\textrm{IIA}: \quad \underbrace{{\mathbb S}^1_{t; r \to 0} \times TN_N^{R\to 0}  \times   {{\bf S}^1} \times {S}^1_{11, r \to 0}}_{\textrm{$k$ D6-branes}}  \times  {\mathbb R^3}.
\label{IIA 5}
\ee
Note however, that we now have a type IIA theory that is strongly-coupled, since the effective type IIA string coupling from a compactification along the circle fibre is proportional to the asymptotic radius which is large (see $\S$2.2 again).

Let us proceed to do a T-duality along ${S}^1_{11}$, which will serve to decompactify the circle, as well as convert the D6-branes to D5-branes in a type IIB theory. By coupling this step with a type IIB S-duality that will convert the D5-branes into NS5-branes, we will arrive at the following type IIB configuration at weak-coupling:
\be
\textrm{IIB}: \quad \underbrace{{\mathbb S}^1_{t; r \to 0} \times TN_N^{R\to 0}  \times  {{\bf S}^1}}_{\textrm{$k$ NS5-branes}}  \times  {\mathbb R^{4}}.
\label{IIB 6}
\ee

Finally, let us do a T-duality along $\mathbb S^1_{t; r\to 0}$, which will bring us back to a type IIA background with NS5-branes and $g^A_s \to \infty$.\footnote{Recall that we have the relation ${\widetilde g}_s = {g_s l_s / r}$, where $\widetilde g$ is the T-dual coupling; that is, if $r \to 0$, $\widetilde g \to \infty$, even though $g_sl_s$ is small.}  Lifting this IIA background back up to M-theory, we will arrive at the following configuration:
\be
\textrm{M-theory}: \quad \underbrace{\mathbb R_t \times TN_N^{R\to 0}  \times  {{\bf S}^1}}_{\textrm{$k$ M5-branes}}  \times  {\mathbb R^{5}}.
\label{M-theory 7}
\ee
\newline
 Hence, from the chain of dualities described above, we conclude that the six-dimensional M-theoretic compactifications given by (\ref{M-theory 1}) and (\ref{M-theory 7}) ought to be physically $\it{equivalent}$.

\newsubsection{Dual Six-Dimensional Compactifications of M-theory with Five-Branes and Orientifold Five-Planes}

To a stack of coincident M5-branes, one can add a five-plane that is intrinsic to M-theory known as the OM5-plane~\cite{hanany}. In other words, we are looking at the following eleven-dimensional configuration (in Euclidean signature):
\be
\textrm{M-theory}: \quad \mathbb R^{5}  \times  \underbrace{\mathbb R^4 / \mathbb Z_k  \times  {{\bf S}^1} \times \mathbb R_t}_{\textrm{$N$ M5-branes/OM5-plane}},
\label{OM-theory 1}
\ee
Unlike the usual Op-planes, the OM5-plane has no (discrete torsion) variants and is thus unique. Its presence will serve to identify opposite points in the spatial directions transverse to its $\mathbb R^4 / \mathbb Z_k  \times  {{\bf S}^1} \times \mathbb R_t$ worldvolume. As such, the gauge symmetries associated with the stack of M5-branes will be modified, much in the same way how Op-planes will effectively identify open-string states with exchanged Chan-Paton indices that connect between Dp-branes, which consequently results in a modification of the effective worldvolume gauge symmetry. An essential fact to note at this point is that the OM5-plane can be interpreted as a $\it{unique}$ ON$5^-_A$-plane in type IIA string theory under a compactification along an ``eleventh circle'' that is transverse to its worldvolume~\cite{hanany}, and as mentioned earlier, the `-' superscript just indicates that its presence will result in an orthogonal gauge symmetry in the type IIA theory, while the `N' just denotes that it can only be associated with NS5-branes. This means that the presence of an OM5-plane will serve to convert an existing gauge symmetry (in a certain regime) of the worldvolume theory on the stack of coincident M5-branes to an orthogonal and $\it{not}$ symplectic type. This fact will be important in $\S$5.

Let us now take the ``eleventh circle'' to be one of the decompactified directions along the $\mathbb R^{5}$ subspace. We then see that (\ref{OM-theory 1}) actually corresponds to the following ten-dimensional type IIA background with $N$ coincident NS5-branes wrapping $\mathbb R^4 / \mathbb Z_k \times {{\bf S}^1} \times \mathbb R_t$ on top of an ON$5^-_A$-plane, where $g^A_sl_s \to \infty$:
\be
\textrm{IIA}: \quad \mathbb R^{4}  \times  \underbrace{\mathbb R^4 / \mathbb Z_k  \times  {{\bf S}^1} \times \mathbb R_t}_{\textrm{$N$ NS5-branes/ON$5^-_A$-plane}}.
\label{OIIA 2}
\ee

Let us next T-dualise along the $\mathbb R_t$ direction of the NS5-branes/ON$5^-_A$-plane configuration. Recall at this point from our discussion in $\S$2.6, that T-dualising along any one of the worldvolume directions of an NS5-brane/ON$5^-$-plane configuration (where the solution is trivial), will bring us back to an NS5-brane/ON$5^-$-plane configuration. Therefore, we will arrive at the following type IIB configuration where $g^B_s \sim 1$:
\be
\textrm{IIB}: \quad \mathbb R^{4}  \times  \underbrace{\mathbb R^4 / \mathbb Z_k  \times  {{\bf S}^1} \times {\mathbb S}^1_{t;  r \to 0}}_{\textrm{$N$ NS5-branes/ON$5^-_B$-plane}}.
\label{OIIB 3}
\ee
In the above, the ON$5^-_B$-plane is the T-dual counterpart of the ON$5^-_A$-plane. It is also the S-dual counterpart of the usual O$5^-$-plane in type IIB theory~\cite{hanany}.

Now, let us T-dualise along a direction that is $\it{transverse}$ to the stack of NS5-branes/ON$5^-_B$-plane. As explained in $\S$2.6, one will end up with a $\it{singular}$ $SN_N$ manifold with no NS5-branes and no ON$5^-_B$-plane. To this end, note that one can view one of the $\mathbb R$'s in $\mathbb R^{4}$ to be a circle of infinite radius. In doing a T-duality along this circle, we arrive at the following type IIA background:
\be
\textrm{IIA}: \quad  SN_N^{R\to 0}  \times  {\mathbb R^4 / \mathbb Z_k}  \times  {{\bf S}^1} \times {\mathbb S}^1_{t; r\to 0},
\label{0IIA 4}
\ee
where $SN_N^{R \to 0}$ is Sen's four-manifold with a $D_{N}$ singularity at the origin and asymptotic radius $R \to 0$. (Recall from $\S$2.6 that $R \to 0$ because we are T-dualising along a trivially fibred circle of infinite radius). This is consistent with the fact that a T-duality along a direction transverse to the ON$5^-_B$-plane gives rise to a solution that can be identified with a unique OM6-plane in M-theory~\cite{hanany}, which, in turn, implies the ${\cal I}_4$ symmetry that is inherent in Sen's four-manifold. At this stage, one also finds that $g^A_s \to 0$. In other words, (\ref{0IIA 4}) can also be interpreted as the following M-theory background with a very small ``eleventh circle'' ${S}^1_{11}$:
\be
\textrm{M-theory}: \quad {\mathbb S}^1_{t; r \to 0} \times SN_N^{R\to 0} \times {{\bf S}^1} \times S^1_{11; r \to 0} \times {\mathbb R^4 / \mathbb Z_k}.
\label{OM-theory 5}
\ee

Recall from $\S$2.1 that the singular ALE space $\mathbb R^4 / \mathbb Z_k$ is simply $TN_k$ with an $A_{k-1}$ singularity at the origin and with asymptotic radius $R \to \infty$. Recall also from $\S$2.2 that M-theory on such a space is equivalent upon compactification along its circle fibre to type IIA string theory with $k$ $\it{coincident}$ D6-branes filling out the directions transverse to this space. In other words, starting from (\ref{OM-theory 5}), one can descend back to the following type IIA background:
\be
\textrm{IIA}: \quad \underbrace{{\mathbb S}^1_{t; r \to 0} \times SN_N^{R\to 0}  \times   {{\bf S}^1} \times {S}^1_{11, r \to 0}}_{\textrm{$k$ D6-branes}}  \times  {\mathbb R^3}.
\label{OIIA 6}
\ee
Note however, that we now have a type IIA theory that is strongly-coupled, since the effective type IIA string coupling from a compactification along the circle fibre is proportional to the asymptotic radius which is large (see $\S$2.2 again).

Let us proceed to do a T-duality along ${S}^1_{11}$, which will serve to decompactify the circle, as well as convert the D6-branes to D5-branes in a type IIB theory. By coupling this step with a type IIB S-duality that will convert the D5-branes into NS5-branes, we will arrive at the following type IIB configuration at weak-coupling:
\be
\textrm{IIB}: \quad \underbrace{{\mathbb S}^1_{t; r \to 0} \times SN_N^{R\to 0}  \times  {{\bf S}^1}}_{\textrm{$k$ NS5-branes}}  \times  {\mathbb R^{4}}.
\label{OIIB 7}
\ee

Finally, let us do a T-duality along $\mathbb S^1_{t; r \to 0}$, which will bring us back to a type IIA background with NS5-branes and $g^A_s \to \infty$.\footnote{Recall that we have the relation ${\widetilde g}_s = {g_s l_s / r}$, where $\widetilde g$ is the T-dual coupling; that is, if $r \to 0$, $\widetilde g \to \infty$, even though $g_sl_s$ is small.} Lifting this IIA background back up to M-theory, we will arrive at the following configuration:
\be
\textrm{M-theory}: \quad \underbrace{SN_N^{R\to 0}  \times  {{\bf S}^1} \times \mathbb R_t}_{\textrm{$k$ M5-branes}}  \times  {\mathbb R^{5}}.
\label{OM-theory 8}
\ee
\newline
 Thus, from the chain of dualities described above, we conclude that the six-dimensional M-theoretic compactifications given by (\ref{OM-theory 1}) and (\ref{OM-theory 8}) ought to be physically $\it{equivalent}$.

\newsection{A Physical Interpretation of a Two-Dimensional Geometric Langlands Duality}

\vspace{-0.8cm}
In this next-to-last section, we will show that the Braverman-Finkelberg (BF) relation~\cite{BF} for $A_{N-1}$ groups can be understood as an invariance in the spacetime BPS spectra of six-dimensional compactifications of M-theory under string dualities. To this end, we shall first review some essential facts about the BF-relation, and how it serves to define a two-dimensional extension to complex surfaces of the usual geometric Langlands duality involving complex curves. Thereafter, with reference to the equivalent six-dimensional compactifications discussed in the previous section, we will proceed to explain the physical interpretation of the BF-relation for $A_{N-1}$ groups. In addition, by making the appropriate modifications to our original setup, we will make contact and find agreement with a closely related result demonstrated earlier via purely field-theoretic considerations by Witten in \cite{Lectures by Witten}.

\newsubsection{Towards a Two-Dimensional Geometric Langlands Duality}

For a group $G$, let $G^{\vee}$ be its Langlands dual. Abstractly speaking, one of the main properties of the geometric Satake isomorphism is that it associates an irreducible $G^{\vee}$-module $L(\tilde\lambda)$ labeled by a dominant weight $\tilde\lambda$ of $G^{\vee}$, with the intersection cohomology $\textrm{IC}(\overline{\textrm{Gr}}^{\tilde\lambda}_G)$ of $\overline{\textrm{Gr}}^{\tilde\lambda}_G$, where $\overline{\textrm{Gr}}^{\tilde\lambda}_G$ is a (usually singular) projective variety related to the affine Grassmannian of $G$ with coweight $\tilde\lambda$. As a byproduct of this isomorphism, one can compute some $\tilde\mu$-graded aspect of $\textrm{IC}(\overline{\textrm{Gr}}^{\tilde\lambda}_G)$, in terms of some $\tilde\mu$-graded aspect of $L(\tilde\lambda)$, where $\tilde\mu$ is a weight of $G^{\vee}$.

In \cite{BF}, Braverman and Finkelberg attempted to extend the above geometric Satake isomorphism to affine Kac-Moody groups $G_{\textrm{aff}}$ and $G^{\vee}_{\textrm{aff}}$. Towards this extension, they constructed $L(\lambda)$, which can be interpreted as an irreducible $G^{\vee}_{\textrm{aff}}$-module of highest dominant weight $\lambda$ of $G^{\vee}_{\textrm{aff}}$, as well as the intersection cohomology $\textrm{IC}(\overline{\textrm{Gr}}^{\lambda}_{G_{\textrm{aff}}})$ of $\overline{\textrm{Gr}}^{\lambda}_{G_{\textrm{aff}}}$, where $\overline{\textrm{Gr}}^{\lambda}_{G_{\textrm{aff}}}$ is a (usually singular) affine quiver variety related to the affine Grassmannian of $G_{\textrm{aff}}$ with coweight $\lambda$. As a byproduct of their extension of the geometric Satake isomorphism, they conjectured that some $\mu$-graded aspect of $\textrm{IC}(\overline{\textrm{Gr}}^{\lambda}_{G_{\textrm{aff}}})$ ought to be computable in terms of some $\mu$-graded aspect of  $L(\lambda)$, where $\mu$ is now a dominant weight of $G^{\vee}_{\textrm{aff}}$. Before we elaborate further on the relation implied by their conjecture, let us first review some essential statements furnished in \cite{BF}.

For any semisimple group $G$ that is simply-connected, the non-compact moduli space $\textrm{Bun}_G(\mathbb R^4/\mathbb Z_k)$ of holomorphic $G$-bundles on ${\mathbb R}^4/{\mathbb Z}_k$, is connected with components labeled by $\mu = (k, \bar\mu, j)$ and $\lambda = (k, \bar\lambda, m)$, where $\lambda$ and $\mu$ are lifts to $G^{\vee}_{\textrm{aff}}$ of the dominant weights $\bar\lambda$ and $\bar\mu$ of $G^{\vee}$ at level $k$, and $j,m \in \mathbb Z$. $\bar \lambda$ and $\bar\mu$ are determined by the conjugacy classes of the homomorphism $\rho:{\mathbb Z}_k \to G$ associated with the ${\mathbb Z}_k$-action in the fibre of the $\mathbb Z_k$-equivariant $G$-bundle on $\mathbb R^4$ (which is the same as the $G$-bundle on $\mathbb R^4 /\mathbb Z_k$) at the origin and infinity, respectively, and the second Chern-class of the bundle is given by $a = k(m-j) + {\bar \lambda}^2 - {\bar \mu}^2$.

The Braverman-Finkelberg relation can then be stated as \cite{BF}
\be
\textrm{dim} V^{\lambda}_{\mu} = \textrm{dim}L(\lambda)_{\mu}.
\label{BF}
\ee
Let us explain the terms that appear above. Firstly, $V^{\lambda}_{\mu}$ is the (global) intersection cohomology of (the Uhlenbeck compactification of) $\textrm{Bun}^{\lambda}_{G,\mu}(\mathbb R^4/\mathbb Z_k)$, the $(\lambda, \mu)$-component of (the usually singular) $\textrm{Bun}_G(\mathbb R^4/\mathbb Z_k)$. Secondly, $L(\lambda)_{\mu}$ is the $\mu$-weight space of $L(\lambda)$, where for the Lie algebra ${\frak g}^{\vee}_{\textrm{aff}}$ of $G^{\vee}_{\textrm{aff}}$, $L(\lambda)$ can also be regarded as an integrable module over ${\frak g}^{\vee}_{\textrm{aff}}$  at level $k$ of highest dominant weight $\lambda$. Moreover, $L(\lambda)_{\mu}$ is known to be finite-dimensional. Note that (\ref{BF}) has been proved for the $A_{N-1}$ groups, i.e., $G=SU(N)$ (see $\S$7 of \cite{BF}). However, it remains to be proved for all other semisimple and simply-connected groups. The Braverman-Finkelberg conjecture can, in this sense, be regarded as an extension to complex surfaces of the usual geometric Langlands duality which involves the moduli space of holomorphic $G$-bundles on complex curves instead.

In preparation of our upcoming physical interpretation of the mathematical relation (\ref{BF}) for $G=SU(N)$, let us now note the following points. Firstly, by the Hitchin-Kobayashi correspondence, (a compactification of) $\textrm{Bun}_G(\mathbb R^4/\mathbb Z_k)$  can be identified with (a compactification of) the moduli space ${\cal M}_G(\mathbb R^4/ \mathbb Z_k)$ of $G$-instantons on ${\mathbb R}^4/{\mathbb Z}_k$, where the instanton number is given by $a$. Secondly, the conjugacy classes of the homomorphism $\rho:\mathbb Z_k \to G$ associated with the $\mathbb Z_k$-action at the origin and infinity, correspond to flat connections of the $G$-bundle at those points. Next, note that the intersection cohomology of (the Uhlenbeck compactification of) ${\cal M}^{\lambda}_{G, \mu}(\mathbb R^4/ \mathbb Z_k)$, can be interpreted as its ${\bf L}^2$-cohomology~\cite{braverman}. Last but not least, note that if the Lie-algebra $\frak g$ of $G$ is simply-laced, then ${\frak g}^{\vee}_{\textrm{aff}} \simeq {\frak g}_{\textrm{aff}}$.

Hence, for $G=SU(N)$, the above points mean that the Braverman-Finkelberg relation will be given by
\be
\textrm{dim}\left[{\textrm H}^\ast_{{\bf L}^2}{\cal U}({\cal M}^{\lambda}_{SU(N), \mu}(\mathbb R^4/ \mathbb Z_k))\right]= \textrm{dim} \left [\widehat{su}(N){}^{\lambda, k}_{\mu}\right],
\label{BF SU(N)}
\ee
where ${\textrm H}^\ast_{{\bf L}^2}{\cal U}({\cal M}^{\lambda}_{SU(N), \mu}(\mathbb R^4/ \mathbb Z_k))$ is the ${\bf L}^2$-cohomology of of the Uhlenbeck compactification ${\cal U}({\cal M}^{\lambda}_{SU(N), \mu}(\mathbb R^4/ \mathbb Z_k))$ of the $(\lambda, \mu)$-component of the moduli space of $SU(N)$-instantons on $\mathbb R^4/\mathbb Z_k$, and $\widehat{su}(N){}^{\lambda, k}_{\mu}$ is the $\mu$-weight space of the integrable module $\widehat{su}(N){}^{\lambda, k}_{\ast}$ over $\frak{su}(N)_{\textrm{aff}}$ at level $k$ of highest dominant weight $\lambda$. Moreover, it can also be shown~\cite{BF} that the compact space ${\cal U}({\cal M}^{\lambda}_{SU(N), \mu}(\mathbb R^4/ \mathbb Z_k))$ is isomorphic to an affine Nakajima quiver variety~\cite{BF 22}, which implies that it can be endowed with a hyperk\"ahler structure.

\newsubsection{A Physical Interpretation of the BF-relation for  $A_{N-1}$ Groups}

We shall now proceed to elucidate the physical intepretation of (\ref{BF SU(N)}). To this end, let us consider the M-theory configuration discussed in $\S$3.1
\be
\textrm{M-theory}: \quad \mathbb R^{5}  \times  \underbrace{\mathbb R^4 / \mathbb Z_k  \times  {{\bf S}^1}\times \mathbb R_t}_{\textrm{$N$ M5-branes}},
\label{M-theory 1 discussion}
\ee
and its $\it{dual}$ M-theory configuration
\be
\textrm{M-theory}: \quad \underbrace{TN_N^{R\to 0}  \times  {{\bf S}^1}\times \mathbb R_t}_{\textrm{$k$ M5-branes}}  \times  {\mathbb R^{5}}.
\label{M-theory 7 discussion}
\ee
Notice that since $\mathbb R^4 / \mathbb Z_k$ and $TN_N^{R\to 0}$ are hyperk\"ahler four-manifolds which break half of the total thirty-two supersymmetries,  the resulting six-dimensional spacetime theories along $\mathbb R^5 \times \mathbb R_t$ in (\ref{M-theory 1 discussion}) and (\ref{M-theory 7 discussion}) will both have $(1,1)$ supersymmetry. As such, one can define spacetime BPS states which are annihilated by part of the $(1,1)$ supersymmetry algebra. Since the supersymmetries of the worldvolume theory on the stack of five-branes come from the background spacetime supersymmetries which are left unbroken in their presence, spacetime BPS states will correspond to worldvolume BPS states living in the stack of five-branes. In particular, in six-dimensional compactifications of M-theory such as that given by (\ref{M-theory 1 discussion}) and (\ref{M-theory 7 discussion}), spacetime BPS states will correspond to half-BPS states of the worlvolume theory of the five-branes~\cite{DVV}. For example, in a six-dimensional compactification of M-theory with a five-brane wrapping $K3 \times {\bf S}^1$, the half-BPS states of the worldvolume theory correspond to the 16-dimensional massless representations of the $(1,1)$ spacetime supersymmetry algebra~\cite{DVV}. For our purpose of elucidating the physical interpretation of (\ref{BF SU(N)}), it suffices to describe the spectrum of these spacetime BPS states in (\ref{M-theory 1 discussion}) and (\ref{M-theory 7 discussion}). In order to do this however, one must first ascertain the quantum worldvolume theory of the stack of five-branes.

\bigskip\noindent{\it{Quantum Worldvolume Theory on the Five-Branes}}

In ten-dimensions or less, the fundamental string and in particular its magnetically dual NS5-brane, have their origins in the two- and five-brane in M-theory, respectively. From this perspective, we see that the five-branes must be as fundamental as the strings themselves. One can then expect that upon a quantisation of the worldvolume theory of five-branes, we would get a spectrum spanned by a tower of excited states, much like that in any other string theory. Indeed, it is known that the quantum worldvolume theory of $l$ coincident five-branes is given by a second-quantised string theory living in the six-dimensional worldvolume that does not contain gravity~\cite{RD}. In the low-energy limit, this string theory reduces to an $A_{l-1}$ $(2,0)$ superconformal field theory of $l-1$ massless tensor multiplets,\footnote{Actually, there are $l$ such tensor multiplets to begin with from the $l$ five-branes. However, a single tensor multiplet, whose scalars describe the transverse position of the centre of mass of the $l$ five-branes, is not really part of the worldvolume theory and has been factored out. This is analogous to the case of $N$ parallel D3-branes in type IIB theory, where the worldvolume theory has $SU(N)$ instead of $U(N) \simeq SU(N) \times U(1)$ gauge symmetry, where the extra $U(1)$ factor corresponding to the centre of mass position of the D3-branes has been factored out.} each consisting of a chiral two-form $Y$ (i.e., with self-dual field strength $dY = \ast dY$), an $Sp(4)$ symplectic Majorana-Weyl fermion $\psi$ and an $SO(5)$ vector $\phi^A$ of scalars that parameterise the transverse position of the five-branes in eleven-dimensions. ($Sp(4) \simeq SO(5)$ is the $R$-symmetry of the $(2,0)$ superconformal algebra.) It is also known~\cite{DVV,RD,RD ref} that one can describe the quantum worldvolume theory as a sigma-model on instanton moduli space - in an appropriate gauge, the second-quantised string theory living in the worldvolume $M \times {\bf S}^1 \times \mathbb R_t$ of the $l$ five-branes, where $M$ is some hyperk\"ahler four-manifold, can be described by a two-dimensional ${\cal N} = (4,4)$ sigma-model on ${\bf S}^1 \times \mathbb R_t$ with target the moduli space ${\cal M}_{SU(l)}(M)$ of $SU(l)$-instantons on $M$. This is consistent with the fact that ${\cal M}_{SU(l)}(M)$ can generically be endowed with a hyperk\"ahler structure also, which leads to ${\cal N} = (4,4)$ supersymmetry in the sigma-model theory.

Note at this point that upon compactification of the (low-energy) worlvolume theory of the five-branes on ${\bf S}^1$, we get a five-dimensional gauge theory with gauge group $SU(l)$, where the five-dimensional gauge coupling is $g_5^2 = R_s$, and $R_s$ is the radius of the ${\bf S}^1$. The five-dimensional theory has, in addition to the gauge bosons, static particle-like BPS configurations which appear as $SU(l)$-instantons on $M$. Their energy is known to be given by $n/R_s$, where $n$ is the instanton number, which then leads to the association of $n$ with the eigenvalue of the momentum operator $L_0$ along the ${\bf S}^1$~\cite{Seiberg ref}. Hence, ${\cal M}_{SU(l)}(M)$ will consist of distinct components labeled by the eigenvalue $n$ of $L_0$. Consequently, the Hilbert space of the quantum worldvolume theory of the five-branes will be $n$-graded and given by
\be
{\cal H} = {\bigoplus}_{n \geq 0} \ q^n {\cal H}_n,
\label{Hilbert space general}
\ee
where ${\cal H}_n$ is the Hilbert space of the sigma-model on the $n^{\textrm{th}}$-component ${\cal M}^n_{SU(l)}(M)$ of ${\cal M}_{SU(l)}(M)$. Note that $q = e^{2 \pi i \tau}$ (where $\tau$ is in the upper half-plane $\mathbb H$) has been formally included in the above expression to make manifest the $L_0$-action common in two-dimensional sigma-models on ${\bf S}^1 \times \mathbb R_t$, which generates rotations along the ${\bf S}^1$.

\bigskip\noindent{\it{Half-BPS States of the Quantum Worldvolume Theory on the Five-Branes}}

The half-BPS states of the worldvolume theory of the five-branes that correspond to the spacetime BPS states we are seeking, come from the low-energy ground states of the second-quantised string theory~\cite{DVV}. The ground states of the string theory correspond to the states  which are annihilated by all eight supercharges of the ${\cal N} = (4,4)$ sigma-model which therefore span its topological sector. As such, they correspond, in the $n^{\textrm{th}}$-sector, to differential forms on the (compactification of the) target space ${\cal M}^n_{SU(l)}(M)$. Note at this point that the eight supercharges of the sigma-model are in one-to-one correspondence with half of the sixteen supercharges of the worldbrane supersymmetry that are left unbroken by the string theory living in the worldvolume of the five-branes itself~\cite{DVV}. Hence, these eight supercharges, being spinors like $\psi$, will transform in a fundamental $\bf{4}$ of $Sp(4)$. Since on a hyperk\"ahler manifold such as (the compactification of) ${\cal M}^n_{SU(l)}(M)$, a differential form which is annihilated by eight operators that transform in the fundamental representation of its $Sp(4)$ ``Lefschetz'' action must be ${\bf L}^2$-harmonic, it will mean that the ground states of the string theory and hence the half-BPS states that we are interested in, must correspond, in the $n^{\textrm{th}}$-sector of the Hilbert space of states, to ${\bf L}^2$-harmonic forms which span the ${\bf L}^2$-cohomology of (the compactification of) ${\cal M}^n_{SU(l)}(M)$.\footnote{It is a theorem that on any complete manifold, an ${\bf L}^2$-harmonic form represents a class in the ${\bf L}^2$-cohomology~\cite{Hitchin}.}

\bigskip\noindent{\it{Spacetime BPS States from $N$ Five-Branes in M-Theory on $\mathbb R^4 / \mathbb Z_k  \times {\bf S}^1$}}

In regard to the M-theory configuration (\ref{M-theory 1 discussion}), the four-manifold $M$ will correspond to $\mathbb R^4/ \mathbb Z_k$, and $l$ will correspond to $N$. Note at this juncture that for the instanton action to be finite even in the presence of the singularity of $M = \mathbb R^4/\mathbb Z_k$ at the origin, we must restrict to flat connections over this point, i.e., we must consider conjugacy classes of the homomorphism $\rho:\mathbb Z_k \to SU(N)$ associated with a $\mathbb Z_k$-action in the fibre of the $\mathbb Z_k$-equivariant $SU(N)$-bundle over $\mathbb R^4$ (which is the same as an $SU(N)$-bundle over $\mathbb R^4/\mathbb Z_k$) at the origin. Moreover, since $M$ is non-compact, in order for the instanton action to be finite in an integration over $M$, we must restrict to flat connections at infinity, i.e., we must consider conjugacy classes of the homomorphism $\rho:\mathbb Z_k \to SU(N)$ associated with a $\mathbb Z_k$-action in the fibre of the $\mathbb Z_k$-equivariant $SU(N)$-bundle over $\mathbb R^4$ at infinity. This means that the Hilbert space of the corresponding sigma-model should also be graded by the different conjugacy classes of the homomorphism $\rho:\mathbb Z_k \to SU(N)$ associated with the origin and infinity of $M = \mathbb R^4/ \mathbb Z_k$.

Indeed, according to our discussion in $\S$4.1, the (Uhlenbeck compactification of the) moduli space ${\cal M}^n_{SU(l)}(M)$ will, in this case, be given by the compact hyperk\"ahler manifold ${\cal U}({\cal M}^{\lambda}_{SU(N), \mu}(\mathbb R^4/ \mathbb Z_k))$, where $n$ will correspond to $a = \{\lambda, \mu\} = k(m-j) + {\bar\lambda}^2 - {\bar \mu}^2$, and where $\bar\lambda$ and $\bar\mu$ are determined by the conjugacy classes of the homomorphism $\rho:\mathbb Z_k \to SU(N)$ associated with the origin and infinity, respectively. (Recall that the pairs $(\lambda, \mu)$, which label the distinct but connected components of the moduli space, are such that they carry the data $\lambda = (k,\bar\lambda, m)$ and $\mu = (k, \bar \mu, j)$, where $m,j \in \mathbb Z$.) In other words, the Hilbert space of the worldvolume theory of the $N$ coincident five-branes on $\mathbb R^4/\mathbb Z_k \times {\bf S}^1 \times \mathbb R_t$ will be given by
\be
{\cal H}_{SU(N)} = {\bigoplus}_{\lambda, \mu} \ q^{\{\lambda, \mu\}} {\cal H}_{\lambda, \mu},
\label{Hilbert space SU(N)}
\ee
where ${\cal H}_{\lambda, \mu}$ is the Hilbert space of the sigma-model on ${\cal U}({\cal M}^{\lambda}_{SU(N), \mu}(\mathbb R^4/ \mathbb Z_k))$. Note that (\ref{Hilbert space SU(N)}) is expressed up to a shift in the $L_0$-action - there exists non-extremal components of ${\cal M}_{SU(N)}(\mathbb R^4/ \mathbb Z_k)$ where the $L_0$-action is shifted by $-kc$, where $c$ is some real number (see Lemma 4.9 of \cite{BF}). As we will soon see, this is consistent with a shift in the overall conformal dimension of the spectrum of states in the highest weight modules of a relevant $SU(N)$ WZW model at level $k$, in the physically dual picture.

Thus, the half-BPS states of the quantum worldvolume theory of the five-branes given by the ground states of the sigma-model, will, in the $(\lambda, \mu)$-sector of the Hilbert space of states, correspond to classes in the ${\bf L}^2$-cohomology of ${\cal U}({\cal M}^{\lambda}_{SU(N), \mu}(\mathbb R^4/ \mathbb Z_k))$. Therefore, one can write the Hilbert space of half-BPS states of the worldvolume theory of the $N$ five-branes in (\ref{M-theory 1 discussion}) as
\be
{\cal H}^{\textrm {BPS}}_{SU(N)} = \bigoplus_{\lambda, \mu} {\textrm H}^\ast_{{\bf L}^2}{\cal U}({\cal M}^{\lambda}_{SU(N), \mu}(\mathbb R^4/ \mathbb Z_k)).
\label{Hilbert space SU(N) explicit}
\ee
  As explained earlier, ${\cal H}^{\textrm {BPS}}_{SU(N)}$ is also the Hilbert space of the corresponding spacetime BPS states in the resulting six-dimensional theory in (\ref{M-theory 1 discussion}).

\bigskip\noindent{\it{Spacetime BPS States from $k$ Five-Branes in M-Theory on $TN^{R\to 0}_N \times {\bf S}^1$}}

Let us now turn our attention to the dual M-theory configuration in (\ref{M-theory 7 discussion}). It seems that one might be able to proceed as before to ascertain the half-BPS states as the low-energy ground states of the second-quantised string theory living in the six-dimensional worldvolume $TN^{R\to 0}_N \times {\bf S}^1 \times \mathbb R_t$. In order to do so, one would need a mathematical description of the moduli space of $SU(N)$-instantons on $TN^{R\to 0}_N$. However, such a description is currently out of reach, at least within the scope of this paper.

Nevertheless, recall that the low-energy limit of the second-quantised string theory corresponds to an $A_{k-1}$ $(2,0)$ superconformal field theory of massless tensor multiplets. Hence, where the ground states of the string theory are concerned, one can take the quantum worldvolume theory to be (super)conformal. Being (super)conformal, it enjoys a state-operator isomorphism - the states obtained from a quantisation of the theory on a five-sphere at infinity correspond to the local operators in the bulk. In this way, the entire spectrum of the theory can be determined by a quantisation of the theory at infinity. This means that in regard to determining the ground states, one could alternatively analyse the worldvolume theory near the boundary without loss of information. Equivalently, this can be understood as a direct consequence of the (super)conformal invariance of the theory - whereby a rescaling of the worldvolume to bring the region near infinity to a finite distance away from the origin would nevertheless leave the theory invariant - such that one can, at the outset, directly analyse the worldvolume theory near the far boundary instead.

Near the boundary at infinity, the ${\bf S}^1_R$ circle fibre of $TN^{R \to 0}_N$ has radius $R \to 0$. In order to make sense of this limit, note that a compactification along the circle fibre would take us down to a type IIA theory whereby the stack of $k$ coincident five-branes would now correspond to a stack of $k$ coincident D4-branes. In addition, recall from $\S$2.2 that since the circle fibration is a monopole bundle over the ${\bf S}^2$ at infinity of charge $N$, we will also have $N$ D6-branes spanning the directions transverse to the $\mathbb R^3$ base. Moreover, since $TN^{R \to 0}_N$ has an $A_{N-1}$ singularity at the origin, the D6-branes will be coincident. In other words, near the boundary, one can analyse the following type IIA system instead:
\be
\textrm{IIA}: \quad \underbrace{ {\mathbb R}^3 \times  {{\bf S}^1} \times {\mathbb R}_t \times {\mathbb R}^5}_{\textrm{I-brane on ${{\bf S}^1} \times {\mathbb R}_t = k\textrm{D4}\cap N \textrm{D6}$}},
\label{equivalent IIA system 1}
\ee
where we have a stack of $k$ coincident D4-branes whose worldvolume is given by $\mathbb R^3 \times {{\bf S}^1} \times {\mathbb R}_t$, and a stack of $N$ coincident D6-branes whose worldvolume is given by ${{\bf S}^1} \times {\mathbb R}_t \times {\mathbb R}^5$, such that the two stacks intersect along ${{\bf S}^1} \times {\mathbb R}_t$ to form a D4-D6 I-brane system.\footnote{A similar D4-D6 I-brane system has also been studied in \cite{Vafa et al} with a broader objective.} A set of D4-branes and D6-branes that intersect along two flat directions is a supersymmetric configuration. In our case, we have (8,0) supersymmetry on the I-brane. This is consistent with the fact that the half-BPS states (living in the D4-branes) which are invariant under eight supersymmetries, must originate from the I-brane theory on ${\bf S}^1 \times \mathbb R_t$ because there are no normalizable fermionic zero-modes to form BPS multiplets on $\mathbb R^3$.\footnote{I wish to thank E.~Witten for clarifications on this point.} As one should be able to do so, one can also see this from the viewpoint of the second-quantised strings living in the worldvolume of the corresponding five-branes as follows. Firstly, notice that the 4-6 open strings which stretch between the D4- and D6-branes come from open M2-branes whose topology is a disc with an ${\bf S}^1_R$ boundary ending on the five-branes. Secondly, the interval filling the disc and hence, the tension of these open M2-branes, goes to zero as the type IIA open strings approach the I-brane and become massless themselves, thus resulting in tensionless second-quantised closed strings of topology ${\bf S}^1_R$ living in the five-brane worldvolume. Thirdly, the $R \to 0$ limit can be viewed as a low-energy limit of these second-quantised closed strings, such that their spectrum will be spanned by the ground states which correspond to the half-BPS states we are interested in. These three points together mean that the half-BPS states must come purely from the field theory associated with the massless 4-6 strings living on the I-brane. Therefore, let us hereon focus on the I-brane theory.

The massless modes of the 4-6 open strings reside entirely in the Ramond sector. On the other hand, all modes in the NS sector are massive. The massless modes are well-known to be chiral fermions on the two-dimensional I-brane~\cite{Vafa ref, Vafa ref 2}. If we have $k$ D4-branes and $N$ D6-branes, the $kN$ complex chiral fermions
\be
\psi_{i, \bar a}(z), \ \psi^{\dagger}_{\bar i, a}(z), \qquad  i = 1, \dots, k, \quad a = 1, \dots, N
\label{4.9}
\ee
will transform in the bifundamental representations $(k, \bar N)$ and $(\bar k, N)$ of $U(k) \times U(N)$. Recall at this point from $\S$2.2  that the asymptotic radius $R$ is given by $g^A_s\sqrt{\alpha'}$. Since we are really studying the system at fixed coupling $g^A_s$, the $R \to 0$ limit can be interpreted as the $\alpha' \to 0$ low-energy limit, consistent with our above description of the second-quantised strings with topology ${\bf S}^1_R$ in this same regime. In this limit, all the massive modes decouple. Consequently, one is just left with the chiral fermions which are necessarily free. Their action is then given by (modulo an overall coupling constant)
\be
I = \int d^2z \ \psi^{\dagger} \bar\partial_{A + A'} \psi,
\label{L}
\ee
where $A$ and $A'$ are the restrictions to the I-brane worldvolume ${\bf S}^1 \times \mathbb R_t$ of the $U(k)$ and $U(N)$ gauge fields living on the D4-branes and D6-branes respectively. In fact, the fermions couple (up to certain discrete identifications under the $\mathbb Z_k$ and $\mathbb Z_N$ centres of $U(k)$ and $U(N)$) to the gauge group~\cite{Vafa et al}
\be
U(1) \times SU(k) \times SU(N),
\label{coupling}
\ee
where the $U(1)$ is the anti-diagonal. This point will be relevant shortly.

Note that the chiral fermions on the I-brane are anomalous. Under a gauge transformation of the $U(k)$ gauge field
\be
\delta A = D \epsilon,
\label{gauge variation U(k)}
\ee
where $\epsilon$ is a position-dependent gauge parameter, the effective action of the fermions transforms as
\be
N \int_{{\bf S}^1 \times \mathbb R_t} {\textrm {Tr}}(\epsilon F_A).
\label{anomaly U(k)}
\ee
Likewise, under a gauge transformation of the $U(N)$ gauge field
\be
\delta A' = D\epsilon',
\label{gauge variation U(N)}
\ee
where $\epsilon'$ is again a position-dependent gauge parameter, the effective action of the fermions transforms as
\be
k \int_{{\bf S}^1 \times \mathbb R_t} {\textrm {Tr}}(\epsilon' F_{A'}).
\label{anomaly U(N)}
\ee

Nevertheless, the overall system which consists of the chiral fermions on the I-brane and the gauge fields in the bulk of the D-branes, is gauge-invariant, as required. This is due to the presence of Chern-Simons coupling terms on the D4 and D6-brane worlvolumes that cancel the above gauge anomalies by the process of anomaly inflow~\cite{Vafa ref 2, nissan}. For example, on the D4-branes, there is a term coupling to the RR two-form field strength ${\tilde H}_2$, and it is given by~\cite{Polchinski}
\be
I_{CS} = \int_{{\bf S}^1 \times \mathbb R_t \times \mathbb R^3} {\tilde H}_2 \wedge CS(A),
\ee
where
\be
CS(A) = \textrm{Tr}(AdA + {2\over 3} A \wedge A \wedge A).
\ee
The presence of the $N$ D6-branes contributes a source term to the equations of motion as:
\be
d{\tilde H}_2 = N \cdot \delta_3( {\cal B} \to X_4),
\label{source term 1}
\ee
where $\delta_3( {\cal B} \to X_4)$ is a delta-function supported Poincar\'e dual three-form, $X_4$ is the worldvolume ${\bf S}^1 \times \mathbb R_t \times \mathbb R^3$ of the D4-branes, and $\cal B$ is the part of the D6-brane worldvolume that intersects $X_4$, i.e., ${\cal B} = {\bf S}^1 \times \mathbb R_t$. Under the gauge transformation (\ref{gauge variation U(k)}), $CS(A)$ will transform as
\be
CS(A) \to CS(A) + d\textrm{Tr}(\epsilon F_A).\ee
Consequently, using (\ref{source term 1}), we find that the corresponding variation in $I_{CS}$ will be given by
\be
\delta I_{CS} = \int_{{\bf S}^1 \times \mathbb R_t \times \mathbb R^3} {\tilde H}_2 \wedge d \textrm{Tr} (\epsilon F_A) = - N \int_{{\bf S}^1 \times \mathbb R_t} \textrm{Tr}(\epsilon F_A),
\ee
which cancels the contribution in (\ref{anomaly U(k)}). Similarly, there is also a term coupling to the RR four-form field strength ${\tilde H}_4$ on the D6-branes, and it is given by~\cite{Polchinski}
\be
I'_{CS} = \int_{{\bf S}^1 \times \mathbb R_t \times \mathbb R^5} {\tilde H}_4 \wedge CS(A').
\ee
The presence of the $k$ D4-branes contributes a source term to the equations of motion as:
\be
d{\tilde H}_4 = k \cdot \delta_5( {\cal C} \to X_6),
\label{source term 2}
\ee
where $\delta_5( {\cal C} \to X_6)$ is a delta-function supported Poincar\'e dual five-form, $X_6$ is the worldvolume ${\bf S}^1 \times \mathbb R_t \times \mathbb R^5$ of the D6-branes, and $\cal C$ is the part of the D4-brane worldvolume that intersects $X_6$, i.e., ${\cal C} = {\bf S}^1 \times \mathbb R_t$. Under the gauge transformation (\ref{gauge variation U(N)}), $CS(A')$ will transform as
\be
CS(A') \to CS(A') + d\textrm{Tr}(\epsilon' F_{A'}).\ee
Consequently, using (\ref{source term 2}), we find that the corresponding variation in $I'_{CS}$ will be given by
\be
\delta I'_{CS} = \int_{{\bf S}^1 \times \mathbb R_t \times \mathbb R^5} {\tilde H}_4 \wedge d \textrm{Tr} (\epsilon' F_{A'}) = - k \int_{{\bf S}^1 \times \mathbb R_t} \textrm{Tr}(\epsilon' F_{A'}),
\label{cancel last}
\ee
which cancels the contribution in (\ref{anomaly U(N)}). Hence, we see that the overall system is gauge-invariant and therefore physically consistent.

The system of $kN$ $\it{complex}$ free fermions has central charge $kN$ and gives a direct realisation of $\widehat{u}(kN)_1$, the integrable modules over $\frak{u}(kN)_{\textrm{aff}}$ at level $\it{one}$~\cite{CFT text}. Consequently, there exists the following affine embedding which preserves conformal invariance:
\be
\widehat{u}(1)_{kN} \otimes \widehat{su}(k)_N \otimes \widehat{su}(N)_k \subset \widehat{u}(kN)_1,
\label{affine embedding}
\ee
where this can be viewed as an affine analog of the gauge symmetry in (\ref{coupling}). What this means is that the total Fock space ${\cal F}^{\otimes kN}$ of the $kN$ free fermions can be expressed as
\be
{\cal F}^{\otimes kN} = \textrm{WZW}_{\widehat{u}(1)_{kN}} \otimes \textrm{WZW}_{\widehat{su}(k)_N} \otimes \textrm{WZW}_{\widehat{su}(N)_k},
\label{Fock}
\ee
where $\textrm{WZW}_{\widehat{u}(1)_{kN}}$,  $\textrm{WZW}_{\widehat{su}(k)_N}$ and $\textrm{WZW}_{\widehat{su}(N)_k}$ are the irreducible integrable modules $\widehat{u}(1)_{kN}$, $\widehat{su}(k)_N$ and $\widehat{su}(N)_k$ that can be realised by the spectra of states of the corresponding $\it{chiral}$ WZW models. Consequently, the partition function of the I-brane theory will be expressed in terms of the characters of $\widehat{u}(1)_{kN}$, $\widehat{su}(k)_N$ and $\widehat{su}(N)_k$.

That the free fermions system on the I-brane can be expressed as a tensor product of chiral WZW models, is physically consistent with the following observation as well. Note that by the process of chiral bosonisation~\cite{Ketov}, one can relate a system of free chiral fermions to a system of free chiral bosons. This implies that the free fermion Lagrangian (\ref{L}) that is gauged to $A + A'$, can be studied in terms of an embedding of chiral bosons in a theory of non-chiral bosons (at the free fermion radius)  gauged to $A + A'$. (This embedding of chiral bosons was considered in \cite{Witten}). In turn, this system can be related to a gauge-anomalous WZW model studied in \cite{gauged WZW} that is only holomorphically coupled to $A + A'$. Indeed, the gauge-anomalous WZW model has exactly the same anomalies (\ref{anomaly U(k)}) and (\ref{anomaly U(N)}) under a variation of the $A$ and $A'$ gauge fields. This however, brings us to the next important point.

Note that ${\cal F}^{\otimes kN}$ is the Fock space of the $kN$ free fermions which have $\it{not}$ been coupled to $A$ and $A'$ yet. Upon coupling to the gauge fields, the characters that appear in the overall partition function of the I-brane theory will be reduced. In a generic situation, the free fermions will couple to the gauge group $U(1) \times SU(k) \times SU(N)$ (see  (\ref{coupling})). In our case however, only the $U(k)$ gauge field living on the D4-branes is dynamical; the $U(N)$ gauge field living on the D6-branes should $\it{not}$ be dynamical as the geometry of $TN^{R \to 0}_N$ is fixed in our description.\footnote{Note that one can impose this condition since the D6-branes are non-compact.} Also, it has been argued in \cite{Vafa et al} that for a multi-Taub-NUT space whose ${\bf S}^1$ fibre has a finite radius at infinity, there can be additional topological configurations of the gauge field - in the form of monopoles that go around the ${\bf S}^1$ at infinity - which render the $U(1)$ gauge field non-dynamical; nonetheless, it is clear that one cannot have such configurations when the radius of the ${\bf S}^1$ at infinity is either infinitely large or zero. Therefore, the free fermions will, in our case, couple dynamically to the gauge group $U(1) \times SU(k)$. Schematically, this means that we are dealing with the following partially gauged CFT
\be
\widehat{u}(kN)_1 / \widehat{u}(1)_{kN} \otimes \widehat{su}(k)_N.
\ee
In particular, the $\widehat{u}(1)_{kN}$ and $\widehat{su}(k)_N$ WZW models will be replaced by the corresponding topological $G/G$ models. Consequently, all characters except those of $\widehat{su}(N)_k$ which appear in the overall partition function of the uncoupled free fermions system on the I-brane, will reduce to constant complex factors $q^{\zeta}$ (where $\zeta$ is a real number) after coupling to the dynamical $SU(k)$ and $U(1)$ gauge fields. As such, the $\it{effective}$ overall partition function of the I-brane theory will only be expressed in terms of the characters of $\widehat{su}(N)_k$ and the $q^\zeta$ factors. For example, in the sector labeled by the highest affine weight $\lambda$, the partition function of the I-brane theory will be given by~\cite{CFT text}
\be
Z_{\lambda}(\tau) = q^{h'_{\lambda} - c/24} \sum_{m} d(m) q^m,
\label{partition function SU(N)}
\ee
 where $h'_{\lambda} = h_{\lambda} + \zeta$ and $h_{\lambda}$ is the conformal dimension of the ground state in the integrable module $\widehat{su}(N)^{\lambda,k}_{\ast}$ with corresponding central charge $c$, and $d(m)$ is the number of states in $\widehat{su}(N)^{\lambda,k}_{\ast}$ which have energy level $m$. Hence, we find that the states of the I-brane theory and hence the sought-after half-BPS states, are counted by the states in an $SU(N)$ WZW model at level $k$. The shift in the overall conformal dimension  by $\zeta$ of the spectrum of states in the module $\widehat{su}(N)^{\lambda,k}_{\ast}$, is consistent with the shift in the $L_0$-action mentioned earlier in our discussion of (\ref{Hilbert space SU(N)}) of the dual compactification.

Finally, note that unitarity of any WZW model requires that its spectrum of states be generated by dominant, highest weight irreducible modules over $\frak{su}(N)_{\textrm{aff,k}}$, i.e., $\frak{su}(N)_{\textrm{aff}}$ at level $k$, such that a generic state in any one such module can be expressed as~\cite{CFT text}
\be
|{\mu}'\rangle = E^{- \alpha}_{-n} \dots E^{-\beta}_{-m} |\lambda\rangle, \qquad \forall~~~n,m \geq 0 ~~ \textrm{and} ~~ \alpha, \beta > 0,
\ee
where the $E^{-\gamma}_{-l}$'s are lowering operators that correspond to the respective modes of the currents of $\frak{su}(N)_{\textrm{aff,k}}$ in a Cartan-Weyl basis that are associated with the complement of the Cartan subalgebra, $|\lambda \rangle$ is a highest weight state associated with a dominant highest affine weight $\lambda$, $\mu' = \lambda -\alpha \dots -\beta$ is an affine weight in the weight system ${\widehat\Omega}_{\lambda}$ of $\widehat{su}(N)^{\lambda,k}_\ast$ which is not necessarily dominant, and $\alpha,\beta$ are positive affine roots. Since at each conformal dimension, there can only be a finite number of affine weights and therefore states, all dominant highest weight modules over $\frak{su}(N)_{\textrm{aff,k}}$ are integrable. In particular, each module labeled by a certain highest affine weight $\lambda$ can be decomposed into a direct sum of finite-dimensional subspaces each spanned by states of the form $|\mu'\rangle$ for $\it{all}$ possible positive affine roots $\alpha, \dots, \beta$. These finite-dimensional subspaces of states are the $\mu'$-weight spaces $\widehat{su}(N){}^{\lambda, k}_{\mu'}$. Note at this point that there is a Weyl-group symmetry on these weight spaces that maps $\mu'$ to a dominant weight $\mu$ in ${\widehat\Omega}_{\lambda}$ which also leaves the weight multiplicities or $d(m)$'s in (\ref{partition function SU(N)}), invariant~\cite{CFT text}. Thus, one can also think of the Hilbert space ${\widehat {\cal H}}^{\textrm{BPS}}_{SU(N)}$ of half-BPS states of the worldvolume theory of the $k$ five-branes in (\ref{M-theory 7 discussion}), as being composed out of sectors $[{\widehat {\cal H}}^{\textrm{BPS}}_{SU(N)}]^\lambda_\mu$ labeled by $(\lambda, \mu)$ - which is consistent with the structure of ${{\cal H}}^{\textrm{BPS}}_{SU(N)}$ in (\ref{Hilbert space SU(N) explicit}) - whereby
\be
\textrm{dim}[{\widehat {\cal H}}^{\textrm{BPS}}_{SU(N)}]^\lambda_\mu = \textrm{dim} \left [\widehat{su}(N){}^{\lambda, k}_{\mu}\right].
\ee
As explained earlier, ${\widehat{\cal H}}^{\textrm {BPS}}_{SU(N)}$ is also the Hilbert space of the corresponding spacetime BPS states in the resulting six-dimensional theory in (\ref{M-theory 7 discussion}).

\bigskip\noindent{\it{A Physical Interpretation of the BF-Relation for $A_{N-1}$ Groups}}

With a concrete description of the sought-after spacetime BPS states at hand, we shall now proceed to explain the physical interpretation of the BF-relation. To this end, note that spacetime BPS states are in general stable over different coupling regimes in a theory; consequently, the spectrum of such states will be invariant under string dualities. For example, the six-dimensional M-theory compactifications with five-branes wrapping $T^4 \times S^1 /\mathbb Z_2$ and $K3 \times S^1$, which, are equivalent under highly non-trivial string/string dualities, have the $\it{same}$ spacetime BPS spectra~\cite{DVV}. Moreover, it has also been established that $\it{a~priori}$ $\it{distinct}$ compactifications of $\it{different}$ string theories, which, are believed to be physically dual, go so far as to possess the same degeneracies in their spacetime BPS spectra at each energy level~\cite{BPS references}.

Coming back to our main point, the above observations therefore indicate that the physical duality of the six-dimensional M-theory compactifications (\ref{M-theory 1 discussion}) and (\ref{M-theory 7 discussion}) will imply that their spacetime BPS spectra are the same, i.e., ${\cal H}^{\textrm{BPS}}_{SU(N)} = {\widehat {\cal H}}^{\textrm{BPS}}_{SU(N)}$. In turn, this means that
\be
\textrm{dim}\left[{\textrm H}^\ast_{{\bf L}^2}{\cal U}({\cal M}^{\lambda}_{SU(N), \mu}(\mathbb R^4/ \mathbb Z_k))\right]= \textrm{dim} \left [\widehat{su}(N){}^{\lambda, k}_{\mu}\right],
\label{BF-relation}
\ee
which, is just (\ref{BF SU(N)}) - the Braverman-Finkelberg relation for $G=SU(N)$! In other words, the Braverman-Finkelberg relation for $A_{N-1}$ groups can be understood as an invariance in the spacetime BPS spectra of six-dimensional M-theory compactifications under string dualities.

\newsubsection{A String-Theoretic Derivation of a Closely Related Field-Theoretic Result}

We shall now proceed to make contact with a closely related result demonstrated earlier via purely field-theoretic considerations by Witten in \cite{Lectures by Witten}.

To this end, let us replace $\mathbb R^4/\mathbb Z_k$ in (\ref{M-theory 1}) with the $\it{smooth}$ multi-Taub-NUT manifold $\widetilde {TN}_k$ with $k$ centres whose circle fibre at infinity has a non-zero and finite radius. By repeating the arguments behind (\ref{M-theory 1})-(\ref{M-theory 7}), we find the following six-dimensional M-theory compactification
\be
\textrm{M-theory}: \quad \mathbb R^{5}  \times  \underbrace{\widetilde {TN}_k  \times  {{\bf S}^1}\times \mathbb R_t}_{\textrm{$N$ M5-branes}},
\label{M-theory 1 Witten discussion}
\ee
to be $\it{dual}$ to the following six-dimensional M-theory compactification
\be
\textrm{M-theory}: \quad \underbrace{TN_N^{R\to 0}  \times  {{\bf S}^1}\times \mathbb R_t}_{\textrm{$k$ non-coincident M5-branes}}  \times  {\mathbb R^{5}}.
\label{M-theory 7 Witten discussion}
\ee
Notice that in contrast to the $\mathbb R^4/\mathbb Z_k$ case - due to the separated ${\vec r}_a$ centres  of the smooth multi-Taub-NUT manifold - the $k$ M5-branes will be $\textrm{\it{non-coincident}}$. This difference will have a non-trivial consequence on the physics as we shall see shortly.

\bigskip\noindent{\it{Spacetime BPS States from $N$ Five-Branes in M-Theory on $\widetilde {TN}_k  \times {\bf S}^1$}}

In order to describe the Hilbert space of spacetime BPS states associated with the half-BPS states of the worldvolume theory of the $N$ five-branes in (\ref{M-theory 1 Witten discussion}), first note that since $\widetilde {TN}_k$ is a hyperk\"ahler manifold like $\mathbb R^4/\mathbb Z_k$, the Gieseker compactification ${\cal G}({\cal M}_{SU(N)}(\widetilde {TN}_k))$ of the moduli space of $SU(N)$-instantons on $\widetilde {TN}_k$ will also inherit a hyperk\"ahler structure, consistent with the ${\cal N} =(4,4)$ supersymmetry of the corresponding sigma-model over it which describes the worldvolume theory of the five-branes. The half-BPS states, being annihilated by all eight supercharges of the sigma-model, will be given by its ground states in the topological sector. As explained in the $\mathbb R^4/\mathbb Z_k$ case, the half-BPS states will therefore correspond to harmonic forms in the ${\bf L}^2$-cohomology of ${\cal G}({\cal M}_{SU(N)}(\widetilde {TN}_k))$.

Secondly, for the instanton action to be finite in an integration over noncompact $\widetilde {TN}_k$, we need to have flat albeit nontrivial connections far away from the origin of $\widetilde {TN}_k$. The hyperk\"ahler metrics on  $\widetilde {TN}_k$ are asymptotic at infinity to $(\mathbb  R^3 \times {\bf S}^1)/ \mathbb Z_k$; because gauge-inequivalent classes of flat connections far away from the origin correspond to conjugacy classes of homomorphisms $\rho_{\infty}$ from the fundamental group at infinity to $G$, and that moreover, conjugacy classes of the homomorphism $\phi: Z_m \to G$ are in one-to-one correspondence with dominant affine weights of $SU(N)_{\textrm{aff}}$ of level $m$ (see Lemma 3.3 and $\S$3.4 of \cite{BF}), we find that distinct choices $\phi$ of flat connections far away from the origin will correspond to distinct dominant affine weights $\mu = (k, \bar \mu, j)$ of $SU(N)_{\textrm{aff}}$ at level k.

Thirdly, recall that in the case of $\mathbb R^4 / \mathbb Z_k$, the $k$ centers coincide with multiplicity $k$ at the origin such that a $\mathbb Z_k$-type singularity develops whence we have a $\mathbb Z_k$-action in the fiber of the $G$-bundle at 0. On the other hand, in the case of $\widetilde {TN}_k$,  we have instead $k$ non-coincident centers of multiplicity 1 each -- in other words, we have instead a $\mathbb Z_1$-action in the fiber of the $G$-bundle over each of the $k$ positions ${\vec p}_m$ of the non-coincident centers. Since this action is given by a conjugacy class of the homomorphism $\rho: Z_1 \to G$, we can associate $k$ distinct dominant affine weights $\lambda^{(m)} = (1, \bar \lambda^{(m)}, i^{(m)})$ of $SU(N)_{\textrm{aff}}$ of level $1$  with the $k$ non-coincident centers, where the $i^{(m)}$'s are \emph{a priori} nonzero. Nonetheless, consistency with the results of $\S$4.2 (where all $k$ centers coincide) constrains the $i^{(m)}$'s to be zero.

Noting that $c_1 = 0$ for $SU(N)$-instantons, we conclude from the above points that ${\cal G}({\cal M}_{SU(N)}(\widetilde {TN}_k))$ will consist of connected components labeled by $\{\lambda^{(1)}, \dots, \lambda^{(k)} \}$, $\mu$, and the instanton number $n \geq 0$. Hence, the Hilbert space of the worldvolume theory of the $N$ five-branes in (\ref{M-theory 1 Witten discussion}) can be written as
\be
{\cal H} = \bigoplus_{{\boldsymbol \lambda}, \mu, n} q^n {\cal H}^{\boldsymbol \lambda}_{\mu, n},
\label{q}
\ee
where ${\boldsymbol \lambda} = \sum_{r=1}^k \lambda^{(r)}  = (k, \bar{{\boldsymbol \lambda}}, 0)$,  ${\cal H}^{\boldsymbol \lambda}_{n, \mu}$ is the Hilbert space of the sigma-model on the $({\boldsymbol \lambda}, \mu, n)$-component ${\cal G}({\cal M}^{{\boldsymbol \lambda}}_{SU(N), \mu, n}(\widetilde {TN}_k))$ of ${\cal G}({\cal M}_{SU(N)}(\widetilde {TN}_k))$. Again, $q$ has been formally included to make manifest the $L_0$-action. In turn, (\ref{q}) means that the corresponding partition function of spacetime BPS states (associated with the half-BPS states of the worldvolume theory) in a certain ${\boldsymbol \lambda}$-sector, will be given by
\be
{Z}^{\textrm{BPS}}_{SU(N)} (q) = \sum_{\mu} \sum_{n \geq 0} q^n  \textrm{dim}\left[{\textrm H}^\ast_{{\bf L}^2}{\cal G}({\cal M}^{{\boldsymbol \lambda} }_{SU(N), \mu, n}(\widetilde {TN}_k))\right].
\label{partition function Witten 1}
\ee

\bigskip\noindent{\it{Spacetime BPS States from $k$ Non-Coincident Five-Branes in M-Theory on $TN^{R \to 0}_N  \times {\bf S}^1$}}

We shall now describe the spacetime BPS states associated to the half-BPS states of the worldvolume theory of the $k$ five-branes in the dual compactification (\ref{M-theory 7 Witten discussion}). Proceeding with the same arguments as before, we find that the half-BPS states will be given by the states of the I-brane theory in the following type IIA configuration:
\be
\textrm{IIA}: \quad \underbrace{ {\mathbb R}^3 \times  {{\bf S}^1} \times {\mathbb R}_t \times {\mathbb R}^5}_{\textrm{I-brane on ${{\bf S}^1} \times {\mathbb R}_t = k \ \textrm{non-coincident D4}\cap N \textrm{D6}$}},
\label{equivalent IIA system Witten}
\ee
where we have a stack of $k$ $\textrm{\it{non-coincident}}$  D4-branes whose worldvolume is given by $\mathbb R^3 \times {{\bf S}^1} \times {\mathbb R}_t$, and a stack of $N$ coincident D6-branes whose worldvolume is given by ${{\bf S}^1} \times {\mathbb R}_t \times {\mathbb R}^5$, such that the two stacks intersect along ${{\bf S}^1} \times {\mathbb R}_t$ to form a D4-D6 I-brane system.

It is useful to note at this point that the analysis surrounding (\ref{4.9})-(\ref{Fock}) has also been carried out for a T-dual D5-D5 I-brane system in \cite{nissan}. In particular, one can also understand the embedding (\ref{affine embedding}) as a splitting into the factors $\frak{u}(1)_{\textrm {aff},kN} \times \frak{su}(k)_{\textrm {aff},N} \times \frak{su}(N)_{\textrm {aff},k}$ of the bilinear currents constructed out of the free fermions which nevertheless preserves the total central charge. According to the T-dual analysis in \cite{nissan} of an I-brane that results from stacks of intersecting D5-branes which are separated, the bilinear currents constructed out of the free chiral fermions living on the I-brane in (\ref{equivalent IIA system Witten}) ought to split into the factors $\frak{u}(1)_{\textrm {aff},kN} \times \frak{u}(1)^{k-1}_{\textrm {aff},N} \times \frak{su}(N)_{\textrm {aff},k} \times [\frak{su}(k)_{\textrm {aff},N} / \frak{u}(1)^{k-1}_{\textrm {aff},N}]$. In other words, the system of $kN$ complex free fermions with central charge $kN$ will, in this case, give a realisation of the total integrable module
\be
\widehat{u}(1)_{kN} \otimes [\widehat{u}(1)_{N}]^{k-1} \otimes \left[\widehat{su}(N)_k \otimes {\widehat{su}(k)_N \over{[\widehat{u}(1)_{N}}]^{k-1}}\right],
\label{affine embedding Witten}
\ee
where one can check that the total central charge of the module is $kN$ as required, even under the exchange $k \leftrightarrow N$.

Note also at this juncture that we have the following equivalence 
of coset realisations~\cite{CFT text}
\be
{\widehat{su}(k)_N \over{[\widehat{u}(1)_{N}}]^{k-1}}  = {[\widehat{su}(N)_1]^{k} \over {\widehat{su}(N)_k}}.
\ee
Together with (\ref{affine embedding Witten}), this means that we are effectively dealing with the following total integrable module
\be
\widehat{u}(1)_{kN} \otimes [\widehat{u}(1)_{N}]^{k-1} \otimes [\widehat{su}(N)_1]^k \subset \widehat{u}(kN)_{1},
\label{effective affine embedding Witten}
\ee
which, as indicated above, is also a consistent affine embedding of the $\widehat{u}(kN)_{1}$ module realised by the $kN$ free fermions that also preserves conformal invariance~\cite{Jurgen's text}, as expected. Hence, as in the previous subsection, this means that the total Fock space $F^{kN}$ of the $\it{uncoupled}$ $kN$ free fermions will, in this instance, be given by
\be
{F}^{\otimes kN} = \textrm{WZW}_{\widehat{u}(1)_{kN}} \otimes \textrm{WZW}_{[\widehat{u}(1)_{N}]^{k-1}} \otimes \textrm{WZW}_{[\widehat{su}(N)_1]^k},
\label{Fock Witten}
\ee
where $\textrm{WZW}_{\widehat{u}(1)_{kN}}$, $\textrm{WZW}_{[\widehat{u}(1)_{N}]^{k-1}}$, and $\textrm{WZW}_{[\widehat{su}(N)_1]^k}$ are the irreducible integrable modules $\widehat{u}(1)_{kN}$, $[\widehat{u}(1)_{N}]^{k-1}$ and $[\widehat{su}(N)_1]^k$ that can be realised by the spectra of states of the corresponding $\it{chiral}$ WZW models. Consequently, the partition function of the uncoupled I-brane theory will be expressed in terms of the (product) of characters of $\widehat{u}(1)_{kN}$, $\widehat{u}(1)_{N}$  and $\widehat{su}(N)_1$.

Next, we must couple the free fermions to the gauge fields which are dynamical. Since the $k$ D4-branes are non-coincident, the free fermions will generically couple to the gauge group $U(1) \times U(1)^{k-1} \times SU(N)$, where the $U(1)^{k-1}$ factor is the Cartan tori of $SU(k)$. As explained earlier, since the radius of the circle fibre of $TN^{R \to 0}_N$ goes to zero at infinity, the free fermions will couple dynamically to the $U(1)$ gauge field. In addition, because the geometry of $TN^{R \to 0}_N$ is fixed in our setup, in contrast to the gauge fields on the D4-branes, the $SU(N)$ gauge field on the $N$ D6-branes should $\it{not}$ be dynamical. Hence, we conclude that the free fermions couple dynamically to the gauge group $U(1) \times U(1)^{k-1}$ only. Schematically, this means that we are dealing with the following partially gauged CFT
\be
{\widehat{u}(1)_{kN} \otimes [\widehat{u}(1)_{N}]^{k-1} \otimes [\widehat{su}(N)_1]^k} \over {\widehat{u}(1)_{kN} \otimes [\widehat{u}(1)_{N}]^{k-1}}.
\ee
In particular, the $\widehat{u}(1)_{kN}$ and $[\widehat{u}(1)_{N}]^{k-1}$ WZW models will be replaced by the corresponding topological $G/G$ models. Consequently, all characters except those of $\widehat{su}(N)_1$ which appear in the overall partition function of the uncoupled free fermions system on the I-brane, will reduce to constant complex factors $q^{\delta}$ (where $\delta$ is a real number) after coupling to the dynamical $U(1)$ and $U(1)^{k-1}$ gauge fields. As such, the $\it{effective}$ overall partition function of the I-brane theory will only be expressed in terms of the product of $k$ characters of $\widehat{su}(N)_1$ and the $q^\delta$ factors. For example, in the sector labeled by $k$ dominant highest affine weights $\{\beta^{(1)}, \beta^{(2)}, \dots, \beta^{(k)} \}$, the partition function of the I-brane theory will be given by~\cite{CFT text}
\begin{eqnarray}
Z_{{\boldsymbol\beta}}(q) & = & q^{\delta}~ \left [ \bigotimes_{i = 1}^k \textrm{Tr}_{\beta^{(i)}} \left(  e^{-2\pi i \sum_j u_j J^j_0} q^{L_0 - c'/24} \right) \right] \nonumber \\
& = & q^{\delta} ~ \left [\bigotimes_{i = 1}^k {\Theta^{\textrm{level 1}}_{\beta^{(i)}}(0, q) \over {\eta(q)^{N-1}}} \right],
\label{partition function Witten 2}
\end{eqnarray}
where ${\boldsymbol\beta} = \sum_{r=1}^k \beta^{(r)}  = (k, \bar{{\boldsymbol \beta}}, 0)$, $\eta(q)$ is the usual Dedekind eta-function, $\Theta^{\textrm{level 1}}_{\beta^{(i)}}(\xi, q)$ is the generalised theta-function associated with the highest weight module over $\frak{su}(N)_{\textrm{aff},1}$ labeled by $\beta^{(i)}$ with central charge $c'=N-1$, and $\xi = \sum_{j} u_j J^j_0$ is set to zero because the scalar fields $u_j$ in the Higgs moduli space of the worldvolume theory of the $N$ $\it{coincident}$ D6-branes must vanish since the $SU(N)$ gauge group is not broken down to its Cartan tori generated by the $J^j_0$ bilinear currents.

Thus, we conclude that the partition function of the corresponding spacetime BPS states of the M-theory compactification (\ref{M-theory 7 Witten discussion}) in the sector labeled by $\boldsymbol\beta$, will be given by $Z_{\boldsymbol\beta}(q)$.

\bigskip\noindent{\it{An Agreement with Witten's Field-Theoretic Result}}

Since $\boldsymbol\lambda$ in $Z^{\textrm{BPS}}_{SU(N)}(q)$ of (\ref{partition function Witten 1}) and $\boldsymbol\beta$ in $Z_{\boldsymbol\beta}(q)$ of (\ref{partition function Witten 2}) are both dominant highest affine weights of $SU(N)_{\rm aff}$ of level $k$, an invariance in the spacetime BPS spectrum under string dualities will imply that we have
\be
\bigotimes_{i = 1}^k {\Theta^{\textrm{level 1}}_{\lambda^{(i)}}(q) \over {\eta(q)^{\textrm{rank}(G)}}} = \sum_\mu \sum_{n \geq 0} q^{n - \hat c/24} ~ \textrm{dim}\left[{\textrm H}^\ast_{{\bf L}^2}{\cal G}({\cal M}^{\boldsymbol\lambda}_{G, \mu, n}(\widetilde {TN}_k))\right],
\label{Witten's relation}
\ee
where $G$ is of $A_{N-1}$ type, ${\Theta^{\textrm{level 1}}_{\lambda^{(i)}}(q) / {\eta(q)^{\textrm{rank}(G)}}}$ is the character of the integrable representation (associated with $\lambda^{(i)}$) of the loop group ${\cal L}G$ at level 1, and $\hat c = kc'$ is the central charge of the affine algebra associated with the l.h.s. of (\ref{Witten's relation}).\footnote{To see how one can have $\hat c = kc' = k(N-1)$, first note that $\hat c/24 = \delta = k(h_\lambda -1/24)$ due to the contribution from the topological $G/G$ models, whereby $h_\lambda$ is the conformal dimension of the ground state of the highest weight module of a chiral $\widehat{u}(1)$ WZW model with highest weight $\lambda$. Next, note that the spectra of this WZW model can also be described by that of a free chiral boson on the I-brane ${\bf S}^1 \times \mathbb R_t$, whereby $h_\lambda = {1\over 2} (nr + mr/2)^2$, such that $m,n \in \mathbb Z_{\geq 0}$ and $r$ is the radius of the ${\bf S}^1$~\cite{CFT text}. Since the radius $r$ can be arbitrary, one can always find a solution to $h_{\lambda} = N/24$ for the required values of $n$ and $m$ for some $r$, which, in turn gives us $\hat c = kc'$.}

Note that (\ref{Witten's relation}) has also been derived by Witten in \cite{Lectures by Witten} via purely field-theoretic considerations; in particular, the relation (\ref{Witten's relation}) can be understood as a consequence of an invariance in the BPS spectrum of states in a six-dimensional $A_{N-1}$ superconformal field theory on $\widetilde {TN}_k \times {\bf S}^1 \times \mathbb R_t$ under different limits of a compactification down to five-dimensions. It is indeed satisfying to know that (\ref{Witten's relation}) - which has been derived and understood from a purely string-theoretic perspective as demonstrated above - can also be obtained through a chiefly field-theoretic analysis of a closely related setup rooted in six-dimensional superconformal field theory. Moreover, this agreement in results serves as yet another testament to the universal robustness of the BPS spectrum under different descriptions of the underlying field and string theories in question.

It is interesting to note at this juncture that prop. 7.5 of \cite{BF} (which is an instance of a level-rank duality) seems to suggest that it might be possible to associate the l.h.s. of (\ref{Witten's relation}) with the affine character $\chi^{\widehat{su}(k)_N}$ of the integrable  module $\widehat{su}(k)_N$. Moreover, since ${\cal G}({\cal M}^{\lambda}_{G, n}(\widetilde {TN}_k)$ can be endowed with a smooth hyperk\"ahler structure, its ${\bf L}^2$-cohomology will coincide with its middle-dimensional cohomology~\cite{Hitchin}. In this sense, (\ref{Witten's relation}) can be viewed as a generalisation of Nakajima's celebrated result~\cite{naka} - which relates an expression similar to the r.h.s. of (\ref{Witten's relation}) that involves the middle-dimensional cohomology of the moduli space of $U(N)$-instantons on the $A_{k-1}$ ALE space, to the affine characters $\chi^{\widehat{su}(k)_N}$ - to $SU(N)$-instantons on the smooth multi-Taub-NUT manifold $\widetilde {TN}_k$. That we persist to have affine characters of $\widehat{su}(k)_N$ even though we are considering $SU(N)$-instantons instead of $U(N)$-instantons, is perhaps due to the fact that the underlying four-manifold is a multi-Taub-NUT space and $\it{not}$ an ALE manifold. Indeed, it has been explained in \cite{vafa/witten} that for $SU(N)$-instantons, the affine-algebra side should be given by ``string functions'' $c^{\Lambda}_{\lambda'}$, where $\Lambda$ is a representation of the loop group of $SU(k)$ at level $N$. Since we are dealing with a multi-Taub-NUT space instead of an ALE manifold, according to the analysis in \cite{Vafa et al} for the case of $U(N)$-instantons, there must also appear on the affine-algebra side characters $\chi^{\widehat{u}_1}$ of the integrable representations of the loop group of $U(1)$. Because one can show~\cite{CFT text} that $\chi^{\widehat{u}_1}$ is equivalent to the theta-function $\Theta_{\lambda'}$,\footnote{Actually, the equivalence holds up to a factor of ${\eta(q)}^{-1}$. However, it seems plausible, even though we do not have a concrete way of proving it, that this extra factor will eventually be accounted for, as we are dealing with $SU(N)$-instantons instead of $U(N)$-instantons.} the affine-algebra side should consist of $c^{\Lambda}_{\lambda'}$'s and $\Theta_{\lambda'}$'s in the combination $\sum_{\lambda'} c^{\Lambda}_{\lambda'}\Theta_{\lambda'}$, whence this gives $ \chi^{\widehat{su}(k)_N}$ as claimed. Admittedly, our arguments in regard to (\ref{Witten's relation}) being a generalisation of Nakajima's result, have been somewhat heuristic thus far. Nevertheless, one should be convinced that (\ref{Witten's relation}) is an interesting and thought-provoking relation which begs further mathematical investigation and verification, although this is beyond the scope of our present paper.

A final point to note is that the field-theoretic analysis in \cite{Lectures by Witten} shows that (\ref{Witten's relation}) ought to hold for the other simply-laced $D_N$ and $E_{6,7,8}$ groups as well. In our string-theoretic setup with five-branes, there is no direct way to realise an $E_{6,7,8}$ type gauge symmetry on their worldvolumes. However, as explained briefly in $\S$3.2, one can realise a $D_N$ type gauge symmetry by adding five-planes to the stacks of five-branes. Satisfying and perhaps worthwhile it may be to repeat the above analysis in the presence of a  five-plane, we shall skip it in favour of brevity and proceed with an even more important analysis that will lead us to a mathematically novel Braverman-Finkelberg type relation for the non-simply-connected $D_N$ groups, next.

\newsection{A Braverman-Finkelberg Type Relation for the Non-Simply-Connected $D_N$ Groups}

In this last section, we will argue in favour of a Braverman-Finkelberg (BF) type relation for the non-simply-connected  $D_{N}$ groups, using an invariance in the spacetime BPS spectra of six-dimensional compactifications of M-theory under string dualities which led us to (\ref{BF-relation}) and (\ref{Witten's relation}), as a physical basis.

To this end, let us consider the M-theory configuration discussed in $\S$3.2
\be
\textrm{M-theory}: \quad \mathbb R^{5}  \times  \underbrace{\mathbb R^4 / \mathbb Z_k  \times  {{\bf S}^1} \times \mathbb R_t}_{\textrm{$N$ M5-branes/OM5-plane}},
\label{SO(2N)-theory 1}
\ee
and its $\it{dual}$ M-theory configuration
\be
\textrm{M-theory}: \quad \underbrace{SN_N^{R\to 0}  \times  {{\bf S}^1} \times \mathbb R_t}_{\textrm{$k$ M5-branes}}  \times  {\mathbb R^{5}}.
\label{SO(2N)-theory 2}
\ee
As in the $A_{N-1}$ case, let us proceed to describe the BPS states of the resulting six-dimensional ${\cal N} = (1,1)$ spacetime theories along $\mathbb R^5 \times \mathbb R_t$ in (\ref{SO(2N)-theory 1}) and (\ref{SO(2N)-theory 2}), which are associated with the worldvolume theories on the stack of $N$ M5-branes/OM5-plane on $\mathbb R^4/\mathbb Z_k \times S^1 \times \mathbb R_t$ and $k$ M5-branes on $SN^{R\to 0}_N \times {\bf S}^1 \times \mathbb R_t$, respectively, where $SN^{R\to 0}_N$ is a hyperk\"ahler four-manifold like $\mathbb R^4/\mathbb Z_k $.

\newsubsection{Spacetime BPS States from $N$ Five-Branes/OM5-plane in M-Theory on $\mathbb R^4/\mathbb Z_k \times S^1$}

Due to the presence of the OM5-plane, the low-energy  worldvolume theory of the $N$ M5-branes/OM5-plane stack in (\ref{SO(2N)-theory 1}) will be given by a six-dimensional $D_N$ $(2,0)$ superconformal field theory of $N$ massless tensor multiplets. Consequently, according to our discussions in the previous section, the $\it{full}$ quantum worldvolume theory will be given by a second-quantised string theory living in the six-dimensional worldvolume $\mathbb R^4/\mathbb Z_k \times S^1 \times \mathbb R_t$, which can be described by an ${\cal N} = (4,4)$ sigma-model on ${\bf S}^1 \times \mathbb R_t$ with target the moduli space ${\cal M}_{SO(2N)}(\mathbb R^4/\mathbb Z_k)$ of $SO(2N)$-instantons on $\mathbb R^4/ \mathbb Z_k$. The instanton number will be given by the eigenvalue $n$ of the momentum operator $L_0$ along the ${\bf S}^1$. In addition, because $\mathbb R^4/\mathbb Z_k$ is non-compact and singular at the origin, in order for the instanton action to be finite, one has to restrict to flat connections over the origin and infinity, i.e., one must consider conjugacy classes $\phi_0$ and $\phi_{\infty}$ of the homomorphism $\rho: \mathbb Z_k \to SO(2N)$ associated with the $\mathbb Z_k$-action in the fibre of the $\mathbb Z_k$-equivariant $SO(2N)$-bundle on $\mathbb R^4$ (which is the same as the $SO(2N)$-bundle on $\mathbb R^4/\mathbb Z_k$) at the origin and infinity, respectively. As such, the Hilbert space of the quantum worldvolume theory of the $N$ M5-branes/OM5-plane system will be divided into sectors labeled by $n$, $\phi_0$ and $\phi_{\infty}$:
\be
{\cal H} = {\bigoplus}_{n \geq 0, \phi_0, \phi_{\infty}} \ q^n {\cal H}_{n, \phi_0, \phi_{\infty}},
\label{Hilbert space SO(2N)}
\ee
where ${\cal H}_{n, \phi_0, \phi_{\infty}}$ is the Hilbert space of the sigma-model with target the $(n, \phi_0, \phi_{\infty})$-component ${\cal M}^{n, \phi_0, \phi_{\infty}}_{SO(2N)}(\mathbb R^4/\mathbb Z_k)$ of ${\cal M}_{SO(2N)}(\mathbb R^4/\mathbb Z_k)$.

Since the addition of an OM5-plane will not modify the supersymmetry algebra of the six-dimensional theory on a stack of $N$ five-branes, the spacetime BPS states will,  as before, correspond to the half-BPS states of the worldvolume theory. These half-BPS states, being annihilated by all eight supercharges of the sigma-model, will be given by its ground states in the topological sector. As such, they will correspond to differential forms on (the compactification of) ${\cal M}^{n, \phi_0, \phi_{\infty}}_{SO(2N)}(\mathbb R^4/\mathbb Z_k)$ in the $(n, \phi_0, \phi_{\infty})$-sector of the Hilbert space of states. Just like in the $A_{N-1}$ theory, spinors will transform in the fundamental $\bf {4}$ of $Sp(4)$, i.e., the supercharges of the sigma-model (which are part of the sixteen supercharges of the worldbrane that are left unbroken in the presence of the second-quantised strings) will transform in a $\bf{4}$ of $Sp(4)$. Hence, since the ${\cal N}=(4,4)$ supersymmetry of the sigma-model implies that one can endow (the compactification of) ${\cal M}^{n, \phi_0, \phi_{\infty}}_{SO(2N)}(\mathbb R^4/\mathbb Z_k)$ with a hyperk\"ahler structure, we find that the half- and thus spacetime BPS states, will again be furnished by harmonic forms in the ${\bf L}^2$-cohomology of (the compactification of) ${\cal M}^{n, \phi_0, \phi_{\infty}}_{SO(2N)}(\mathbb R^4/\mathbb Z_k)$.  In summary, the corresponding Hilbert space of spacetime BPS states in the resulting six-dimensional theory in (\ref{SO(2N)-theory 1}) will be given by
\be
{\cal H}^{\textrm {BPS}}_{SO(2N)} = \bigoplus_{n, \phi_{0},\phi_{\infty}} {\textrm H}^\ast_{{\bf L}^2}{\cal U}({\cal M}^{n, \phi_{0},\phi_{\infty}}_{SO(2N)}(\mathbb R^4/ \mathbb Z_k)),
\label{Hilbert space SO(2N) - 1}
\ee
where ${\cal U}({\cal M}^{n, \phi_{0},\phi_{\infty}}_{SO(2N)}(\mathbb R^4/ \mathbb Z_k)$ is an Uhlenbeck compactification of ${\cal M}^{n, \phi_{0},\phi_{\infty}}_{SO(2N)}(\mathbb R^4/ \mathbb Z_k)$.

\newsubsection{Spacetime BPS States from $k$ Five-Branes in M-Theory on $SN^{R \to 0}_N \times {\bf S}^1$}

Let us now turn our attention to the dual compactification (\ref{SO(2N)-theory 2}). One could proceed as above to describe the spacetime BPS states associated with the quantum worldvolume theory of the five-branes on $SN^{R \to 0}_N \times {\bf S}^1 \times \mathbb R_t$. However, one needs a description of the moduli space of instantons on $SN^{R\to 0}_N$, which is currently out of reach, at least within the scope of this paper.

Nonetheless, since the ground states of the second-quantised string - which correspond to the spacetime BPS states - span the low-energy spectrum, it suffices, in regard to these states, to consider the low-energy limit of the string theory described by the $D_N$ $(2,0)$ superconformal field theory of massless tensor multiplets. As explained in the previous section, the conformal invariance of this theory implies that one could alternatively analyse the worldvolume theory near the boundary without loss of information.

 Near the boundary at infinity, the ${\bf S}^1_R$ circle fibre of $SN^{R \to 0}_N$ has radius $R \to 0$. In order to make sense of this limit, note that a compactification along the circle fibre would take us down to a type IIA theory whereby the stack of $k$ coincident five-branes would now correspond to a stack of $k$ coincident D4-branes. In addition, recall from $\S$2.5 that we will also have $N$ D6-branes and an O$6^-$-plane spanning the directions transverse to the $\mathbb R^3/ {\cal I}_3$ base, where ${\cal I}_3$ acts as ${\vec r} \to -{\vec r}$ in $\mathbb R^3$. Moreover, since $SN^{R \to 0}_N$ has a $D_{N}$ singularity at the origin, the D6-branes will be coincident. In other words, near the boundary, one can analyse the following type IIA system instead:
\be
\textrm{IIA}: \quad \underbrace{ {\mathbb R}^3/{\cal I}_3 \times  {{\bf S}^1} \times {\mathbb R}_t \times {\mathbb R}^5}_{\textrm{I-brane on ${{\bf S}^1} \times {\mathbb R}_t = k\textrm{D4}\cap N \textrm{D6}/O$6$^-$}},
\label{equivalent IIA system 2}
\ee
where we have a stack of $k$ coincident D4-branes whose worldvolume is given by $\mathbb R^3/{\cal I}_3 \times {{\bf S}^1} \times {\mathbb R}_t$, and a stack of $N$ coincident D6-branes on top of an O$6^-$-plane whose worldvolume is given by ${{\bf S}^1} \times {\mathbb R}_t \times {\mathbb R}^5$, such that the two stacks intersect along ${{\bf S}^1} \times {\mathbb R}_t$ to form a D4-D6/O$6^-$ I-brane system.

The proceeding analysis of this system is almost identical to the one before for the system (\ref{equivalent IIA system 1}). In particular, the sought-after half-BPS (i.e.~ground) states (living in the D4-branes) will correspond to states of the I-brane theory on ${\bf S}^1 \times \mathbb R_t$. Moreover, the I-brane theory will be a theory of free chiral fermions. As before, the chiral fermions will couple to certain gauge fields. In order to determine what these gauge fields are, let us now discuss what gauge groups should appear in the D4-D6/O$6^-$ I-brane system.

By a T-duality along three directions, we can get to a D1-D9/O$9^-$ system, where O$9^-$ is a spacetime-filling orientifold nine-plane. One can compare this to an analogous D5-D9-O$9^{\pm}$ system studied in \cite{Gimon}, where the gauge groups are different on the D5- and D9-branes; they are either orthogonal on the D5-branes and symplectic on the D9-branes or vice-versa, depending on the sign in O$9^{\pm}$. This is due to the fact that there are four possible mixed Neumann-Dirichlet boundary conditions for the 5-9 open strings which stretch between the corresponding D-branes. On the other hand, there are eight possible mixed Neumann-Dirichlet boundary conditions for the 1-9 open strings stretched between D-branes in the D1-D9/O$9^-$ system, which consequently leads to the $\it{same}$ orthogonal gauge groups on both the D1- and D9-branes. By T-dualising back to a D4-D6/O$6^-$ system, one can conclude that generically, there ought to be, in the presence of the O$6^-$-plane, an $SO(\alpha)$ and $SO(2N)$ gauge group on the $k$ D4- and $N$ D6-branes, respectively, where $\alpha$ depends on $k$. 

To ascertain what $\alpha$ is, note that according to~\cite{nissan}, the total central charge of the \emph{real} chiral fermions   should not change as we move the D4- and D6-branes around; in particular, it should not change as we move the stack of coincident D4- and D6-branes away from the O$6^-$-plane. When we move the stack of coincident D4- and D6-branes away from the O$6^-$-plane, we effectively have the $U(k) \times U(N)$ theory described by (\ref{4.9})--(\ref{L}). Thus, $\alpha$ must be such that the total central charge of the real chiral fermions is $kN$.

Since a single real chiral fermion will contribute one-half to the central charge, we ought to have a total of $2kN$ real chiral fermions. Thus, the massless modes of the 4-6 open strings of the D4-D6/O$6^-$ system will correspond to $2kN$ real chiral fermions that necessarily transform in the bifundamental representation $(\alpha, 2N)$ of $SO(\alpha) \times SO(2N)$ such that $2\alpha N = 2 kN$, i.e., $\alpha = k$. In other words, we would have  $2kN$ real chiral fermions 
\be
\psi_{i, a}(z), \qquad  i = 1, \dots, k, \quad a = 1, \dots, 2N,
\ee
on the I-brane, that will transform in the bifundamental representation $(k, 2N)$ of $SO(k) \times SO(2N)$. From the relation $R = g_s^A \sqrt{\alpha'}$ in $\S$2.5, and the fact that we are studying the system at fixed coupling $g^A_s$, we see that the $R \to 0$ limit can be interpreted as the $\alpha' \to 0$ low-energy limit. In this limit, all the massive modes decouple. Consequently, one is just left with the chiral fermions which are necessarily free. Their action is then given by (modulo an overall coupling constant)
\be
I = \int d^2z \ \psi \bar\partial_{{\cal A} + {\cal A}'} \psi,
\label{L'}
\ee
where $\cal A$ and ${\cal A}'$ are the restrictions to the I-brane worldvolume ${\bf S}^1 \times \mathbb R_t$ of the $SO(k)$ and $SO(2N)$ gauge fields living on the D4-branes and D6-branes in the presence of an O$6^-$-plane, respectively. In other words, the fermions couple to the gauge group
\be
SO(k) \times SO(2N).
\label{coupling'}
\ee
Repeating the calculations in (\ref{gauge variation U(k)})-(\ref{cancel last}), we find that the I-brane theory is anomalous under the corresponding gauge transformations, but the overall D4-D6/O$6^-$ system is anomaly-free and thus physically consistent due to an anomaly-inflow mechanism, just like in the earlier D4-D6 system.

The system of $2kN$ $\it{real}$ free fermions has central charge $kN$ and gives a direct realisation of $\widehat{so}(2kN)_1$, the integrable modules over $\frak{so}(2kN)_{\textrm{aff}}$ at level $\it{one}$~\cite{CFT text}. Consequently, there exists the following affine embedding which preserves conformal invariance~\cite{Hasegawa}:
\be
\widehat{so}(k)_{2N} \otimes \widehat{so}(2N)_{k} \subset \widehat{so}(2kN)_1,
\ee
where this can be viewed as an affine analog of the gauge symmetry in (\ref{coupling'}). What this means is that the total Fock space ${\cal F}^{\otimes 2kN}$ of the $2kN$ free fermions can be expressed as
\be
{\cal F}^{\otimes 2kN} = \textrm{WZW}_{\widehat{so}(k)_{2N}} \otimes \textrm{WZW}_{\widehat{so}(2N)_{k}},
\label{Fock SO(2N)}
\ee
where $\textrm{WZW}_{\widehat{so}(k)_{2N}}$ and $\textrm{WZW}_{\widehat{so}(2N)_{k}}$ are the irreducible integrable modules $\widehat{so}(k)_{2N}$ and $\widehat{so}(2N)_{k}$ that can be realised by the spectra of states of the corresponding $\it{chiral}$ WZW models. Consequently, the partition function of the I-brane theory will be expressed in terms of the characters of $\widehat{so}(k)_{2N}$ and $\widehat{so}(2N)_{k}$.

Note that ${\cal F}^{\otimes 2kN}$ is the Fock space of the $2kN$ free fermions which have $\it{not}$ been coupled to $\cal A$ and ${\cal A}'$ yet. As we saw earlier, upon coupling to the gauge fields, the characters that appear in the overall partition function of the uncoupled I-brane theory will be reduced. In a generic situation, the free fermions will couple to the gauge group $SO(k) \times SO(2N)$ (see  (\ref{coupling'})). In the case at hand, only the $SO(k)$ gauge field living on the D4-branes is dynamical; the $SO(2N)$ gauge field living on the D6-branes/O$6^-$-plane should $\it{not}$ be dynamical as we want the geometry of $SN^{R \to 0}_N$ to be fixed in our description.\footnote{Note that one can impose this condition since the D6-branes/O$6^-$-plane configuration is non-compact.} Therefore, the free fermions will, in this case, couple dynamically to the gauge group $SO(k)$ only. Schematically, this means that we are dealing with the following partially gauged CFT
\be
\widehat{so}(2kN)_1 / \widehat{so}(k)_{2N}.
\ee
In particular, the $\widehat{so}(k)_{2N}$ WZW model will be replaced by the corresponding topological $G/G$ model. Consequently, the characters of $\widehat{so}(k)_{2N}$ which appear in the overall partition function of the uncoupled free fermions system on the I-brane, will reduce to constant complex factors $q^{\zeta'}$ (where $\zeta'$ is a real number) after coupling to the dynamical $SO(k)$ gauge fields. As such, the $\it{effective}$ overall partition function of the I-brane theory will only be expressed in terms of the characters of $\widehat{so}(2N)_{k}$ and the $q^{\zeta'}$ factors. Going through the same arguments as before, we find that the half-BPS states will be counted by the spectrum of states of an $SO(2N)$ WZW model at level $k$. In particular, this means that the Hilbert space ${\widehat {\cal H}}^{\textrm{BPS}}_{SO(2N)}$ of half-BPS states of the worldvolume theory of the $k$ five-branes in (\ref{SO(2N)-theory 2}) can be decomposed into sectors $[{\widehat {\cal H}}^{\textrm{BPS}}_{SO(2N)}]^\lambda_\mu$ labeled by $(\lambda, \mu)$, such that
\be
\textrm{dim}[{\widehat {\cal H}}^{\textrm{BPS}}_{SO(2N)}]^\lambda_\mu = \textrm{dim} \left [\widehat{so}(2N){}^{\lambda, k}_{\mu}\right],
\ee
where $\lambda$ is a highest dominant affine weight, while $\mu$ is a dominant affine weight in the weight system ${\widehat\Omega}_{\lambda}$ of $\widehat{so}(2N){}^{\lambda, k}_{\ast}$. Moreover, ${\widehat{\cal H}}^{\textrm {BPS}}_{SO(2N)}$ is also the Hilbert space of the corresponding spacetime BPS states in the resulting six-dimensional theory in (\ref{SO(2N)-theory 2}).

\newsubsection{A Braverman-Finkelberg Type Relation for the Non-Simply-Connected $D_{N}$ Groups}

In accordance with our arguments in the previous section leading up to (\ref{BF-relation}) and (\ref{Witten's relation}), the physical duality of the six-dimensional M-theory compactifications (\ref{SO(2N)-theory 1}) and (\ref{SO(2N)-theory 2}) will imply that their BPS spectra are the same, i.e., ${\cal H}^{\textrm {BPS}}_{SO(2N)} = {\widehat{\cal H}}^{\textrm {BPS}}_{SO(2N)}$. This in turn means that one should be able to relate the triple $(n, \phi_{0}, \phi_{\infty})$ to the double $(\lambda, \mu)$ such that ${\cal M}^{n, \phi_{0},\phi_{\infty}}_{SO(2N)}(\mathbb R^4/ \mathbb Z_k)$ can be relabeled as ${\cal M}^{\lambda}_{SO(2N), \mu}(\mathbb R^4/ \mathbb Z_k)$, whereby
\be
\textrm{dim} \left[{\textrm H}^\ast_{{\bf L}^2}{\cal U}({\cal M}^{\lambda}_{SO(2N), \mu}(\mathbb R^4/ \mathbb Z_k))\right] = \textrm{dim} \left[\widehat{so}(2N){}^{\lambda, k}_{\mu}\right].
\label{BF-relation for SO(2N)}
\ee
Since the Lie algebra of $D_N$ groups is simply-laced, i.e., $\frak{so}(2N)_{\textrm{aff}} \simeq \frak{so}(2N)^\vee_{\textrm{aff}}$, the above relation can be interpreted as a BF-relation for $D_{N}$ groups!


\begin{thebibliography}{99}

\bibitem{BF}

A.~Braverman, M.~Finkelberg, ``Pursuing the Double Affine Grassmannian I: Transversal Slices via Instantons on $A_{k-1}$ Singularities'', Duke Math. J. Volume {\bf 152}, Number 2 (2010), 175-206, [arXiv:math/0711.2083].

\bibitem{Lectures by Witten}

E.~Witten, ``Duality from Six-Dimensions I, II, III'', lectures delivered at the IAS in Feb 08. Notes for the lectures taken by D.~Ben-Zvi can be found at: [http://www.math.utexas.edu/users/benzvi/GRASP/lectures/IASterm.html]

\bibitem{naka}

H.~Nakajima, ``Instantons on ALE Spaces, Quiver Varieties, and Kac-Moody Algebras'', Duke Math. 76 (1994) 365-416.


\bibitem{BBS}

K.~Becker, M.~Becker, J.H~Schwarz. ``String Theory and M-theory: A Modern Introduction''. (Cambridge Monographs on Mathematical Physics), Cambridge University Press, New York, USA, (2007). 

\bibitem{sen}

A.~Sen, ``A Note on Enhanced Gauge Symmetries in M- and String-Theory'', JHEP 9709 (1997) 001,  [arXiv:hep-th/9707123]

\bibitem{Tong}

D.~Tong, ``NS5-branes, T-duality and Worldsheet Instantons'', JHEP 0207 (2002) 013, [arXiv:hep-th/0204186]

\bibitem{CJ}

C.~Johnson, ``D-branes'', (Cambridge Monographs on Mathematical Physics), Cambridge University Press, New York, USA, (2003). 


\bibitem{sen18}

M.~Atiyah and N.~Hitchin, ``Low energy scattering of nonabelian monopoles'', Phys. Lett. 107A (1985) 21; Phil. Trans. R. Soc. Lond.
A315 (1985) 459; ``The Geometry and Dynamics of Magnetic Monopoles'', Princeton
Univ. Press (1988)

\bibitem{sen17}

N. Seiberg and E. Witten, ``Gauge Dynamics And Compactification To Three Dimensions'', [arXiv:hep-th/9607163]; N. Seiberg, ``IR Dynamics on Branes and Space-Time Geometry'', Phys. Lett. B384 (1996) 81 [arXiv:hep-th/9606017].


\bibitem{hanany}

A.~Hanany, B.~Kol, ``On Orientifolds, Discrete Torsion, Branes and M Theory'', JHEP 0006 (2000) 013 [arXiv:hep-th/0003025]


\bibitem{braverman}

A. Braverman, private communication; M. Goresky, ``${\bf L}^2$-cohomology is Intersection Cohomology'', [http://www.math.ias.edu/~goresky/pdf/zucker.pdf].

\bibitem{BF 22}

H. Nakajima, ``Geometric Construction of Representations of Affine Algebras'', Proceedings of the International
Congress of Mathematicians, Vol. I (Beijing, 2002), 423-438.

\bibitem{DVV}

R.~Dijkgraaf, E.~Verlinde, H.~Verlinde, ``BPS Quantisation of the Five-Brane'', Nucl. Phys.~B486: 89-113, 1997, [arXiv:hep-th/9604055]




\bibitem{RD}

R.~Dijkgraaf, ``The Mathematics of Fivebranes''. International Congress of Mathematicians (ICM 98), Berlin, Germany, Doc. Math. J. DMV, 1999.[arXiv:hep-th/9810157]

\bibitem{RD ref}

J. Schwarz, ``Self-Dual String in Six-Dimensions'', [arXiv:hep-th/9604171]; R. Dijkgraaf, E. Verlinde, H. Verlinde, ``BPS Spectrum of the Five-Brane and Black Hole Entropy'', Nucl. Phys.~B486: 77-88,1997, [arXiv:hep-th/9603126]; R. Dijkgraaf, E. Verlinde, and H. Verlinde, ``5D Black Holes and Matrix Strings'', Nucl. Phys. B506 (1997) 121–142. [arXiv:hep-th/9704018]; O. Aharony, M. Berkooz, S. Kachru, N. Seiberg, and E. Silverstein, ``Matrix Description of Interacting Theories in Six Dimensions'', Phys. Lett. B420 (1998) 55–63, [arXiv:hep-th/9707079]; E. Witten, ``On the Conformal Field Theory of the Higgs Branch'', J. High Energy Phys. 07 (1997) 3, [arXiv:hep-th/9707093].


\bibitem{Seiberg ref}

E. Witten, ``String Theory Dynamics in Various Dimension'', Nucl. Phys. B443
(1995) 85, [arXiv:hep-th/9503124]; M. Douglas, D. Kabat, P. Pouliot, and S. Shenker, ``D-branes and Short Distances in
String Theory'', Nucl. Phys. B485 (1997) 85, [arXiv:hep-th/9608024]; M. Rozali, ``Matrix Theory and U-Duality in Seven Dimensions'', Phys. Lett.~B400 (1997) 260-264, [arXiv:hep-th/9702136].

\bibitem{Hitchin}

N.~Hitchin, ``${\bf L}^2$-Cohomology of Hyperk\"ahler Quotients'', Commun. Math. Phys.  211 (2000) 153-165, [arXiv:math/9909002].

\bibitem{Vafa et al}

R.~Dijkgraaf, L.~Hollands, P.~Sulkowski, C.~Vafa, ``Supersymmetric Gauge Theories, Intersecting Branes and Free Fermions'', JHEP 02 (2008) 106, [arXiv:0709.4446].

\bibitem{Vafa ref}

C. Bachas, M. Green, A. Schwimmer, ``$(8,0)$ Quantum mechanics and symmetry enhancement in type I' superstrings'', JHEP 9801 (1998) 006, [arXiv:hep-th/9712086]; L. Hung, ``Comments on I1-branes'', JHEP 05 (2007) 076,  [arXiv:hep-th/0612207].


\bibitem{Vafa ref 2}

M. Green, J. Harvey, G. Moore, ``I-brane Inflow and Anomalous Couplings on D-branes'',
Class. Quant. Grav. 14 (1997) 47-52, [arXiv:hep-th/9605033].


\bibitem{nissan}

N. Itzhaki, D. Kutasov, N. Seiberg, ``I-brane Dynamics'', JHEP 0601 (2006) 119, [arXiv:hepth/0508025].


\bibitem{Polchinski}

J.~Polchinski, ``String Theory Vol 2: Superstring Theory and Beyond'', (Cambridge
Monographs on Mathematical Physics), Cambridge University Press, New York, USA, (2003). 


\bibitem{CFT text}

P. Di Francesco, P. Mathieu and D. Senechal, ``Conformal Field Theory'', Springer-Verlag, New York, USA, (1999).


\bibitem{Ketov}

S.V.~Ketov, ``Conformal Field Theory'', World Scientific Press, Singapore, (1997). 


\bibitem{Witten}

E.~Witten, ``Five-brane Effective Action in M-theory'', J.Geom.Phys. 22 (1997) 103-133, [arXiv:hep-th/9610234].

\bibitem{gauged WZW}

E.~Witten, ``On Holomorphic Factorization of WZW and Coset Models'', Comm. Math. Phys. 144 (1992), 189-212.



\bibitem{BPS references}

C. Vafa, ``Gas of D-Branes and Hagedorn Density of BPS States'', Nucl. Phys. B463 (1996) 415-419, [arXiv:hep-th/9511026];
``Instantons on D-branes'', Nucl. Phys. B463 (1996) 435-442, [arXiv:hep-th/9512078];
M. Bershadsky, V. Sadov, and C. Vafa, ``D-Branes and Topological Field Theories'', Nucl. Phys. B463 (1996) 420-434, [arXiv:hep-th/9511222];
A. Sen, ``A Note on Marginally Stable Bound States in Type II String Theory'', Phys. Rev. D54: 2964-2967, (1996) [arXiv:hepth/9510229]; ``U Duality and Intersecting D-Branes'', Phys. Rev. D 53: 2874ÐR2877 (1996), [arXiv:hep-th/9511026]; ``T-Duality of
p-Branes'', Mod. Phys. Lett. A11: 827-834 (1996), [arXiv:hep-th/9512062].


\bibitem{Sergey}

Sergey~A.~Cherkis, ``Moduli Spaces of Instantons on the Taub-NUT Space'', Commun. Math. Phys. 290:719-736, 2009, [arXiv:hep-th/0805.1245]

\bibitem{Jurgen's text}

J.~Fuchs, private communication; J.~Fuchs, ``Affine Lie Algebras and Quantum Groups'', (Cambridge Monographs on Mathematical Physics), Cambridge University Press, New York, USA, (1995).




\bibitem{vafa/witten}

C.~Vafa, E.~Witten, `` A Strong Coupling Test of S-Duality'', Nucl.Phys. B431 (1994) 3-77, [arXiv:hep-th/9408074].


\bibitem{Gimon}

E. Gimon, J. Polchinski, ``Consistency Conditions for Orientifolds and D-manifolds'', Phys. Rev. D 54: 1667Ð1676 (1996), 
[arXiv:hep-th/9601038].

\bibitem{Hasegawa}

K. Hasegawa, ``Spin Module Versions of Weyl's Reciprocity Theorem for Classical Kac-
Moody Lie Algebras - An Application to Branching Rule Duality'', RIMS, Kyoto Univ. 25
(1989) 741-828.



\end{thebibliography}
\end{document}